\newcommand{\ab}{\mathbf{a}}

\newcommand{\fb}{\mathbf{f}}

\newcommand{\nb}{\mathbf{n}}

\newcommand{\Tb}{\mathbf{T}}

\newcommand{\sigmab}{\boldsymbol{\sigma}}

\documentclass[11pt]{article}
\usepackage{standalone}
\pdfoutput=1 
\usepackage{multicol}
\usepackage{fullpage}
\usepackage{authblk}
\usepackage{mathtools}
\usepackage{graphicx}
\usepackage{epstopdf}
\usepackage{amsmath}
\usepackage{float}
\usepackage{wrapfig}
\usepackage[normalem]{ulem}
\usepackage[backend=bibtex, style=nature]{biblatex}
\usepackage[usenames, dvipsnames]{color}
\usepackage{xr-hyper}
\usepackage{hyperref}
\newcommand\D{\mathrm{d}}

\definecolor{myRed}{rgb}{0.8500, 0.3250, 0.0980}
\definecolor{myBlue}{rgb}{0, 0.4470, 0.7410}
\definecolor{myGreen}{rgb}{0.5023, 0.7491, 0.4776}
\definecolor{myYellow}{rgb}{0.9139, 0.7258, 0.3063}

\begin{document}
\title{Membrane tension is a key determinant of bud morphology in clathrin-mediated endocytosis}
\author[1]{Julian E. Hassinger}
\author[2]{George Oster}
\author[2]{David G. Drubin}
\author[3]{Padmini Rangamani}
\affil[1]{Biophysics Graduate Group, University of California, Berkeley}
\affil[2]{Department of Molecular and Cell Biology, University of California, Berkeley}
\affil[3]{Department of Mechanical and Aerospace Engineering, University of California, San Diego}

\date{}
\maketitle

\abstract{In clathrin-mediated endocytosis (CME), clathrin and various adaptor proteins coat a patch of the plasma membrane, which is reshaped to form a budded vesicle. Experimental studies have demonstrated that elevated membrane tension can inhibit bud formation by a clathrin coat. In this study, we investigate the impact of membrane tension on the mechanics of membrane budding by simulating clathrin coats that either grow in area or progressively induce greater curvature. At low membrane tension, progressively increasing the area of a curvature-generating coat causes the membrane to smoothly evolve from a flat to budded morphology, whereas the membrane remains essentially flat at high membrane tensions. Interestingly, at physiologically relevant, intermediate membrane tensions, the shape evolution of the membrane undergoes a \emph{snapthrough instability} in which increasing coat area causes the membrane to ``snap'' from an open, U-shaped bud to a closed, $\Omega$-shaped bud. This instability is accompanied by a large energy barrier, which could  cause a developing endocytic pit to stall if the binding energy of additional coat is insufficient to overcome this barrier. Similar results were found for a coat of constant area in which the spontaneous curvature progressively increases. Additionally, a pulling force on the bud, simulating a force from actin polymerization, is sufficient to drive a transition from an open to closed bud, overcoming the energy barrier opposing this transition.}

%\section*{Significance statement}

%Plasma membrane tension plays an important role in various biological processes. In particular, recent experimental studies have shown that membrane tension opposes membrane remodeling processes such as clathrin-mediated endocytosis (CME). We have identified a mathematical relationship between the curvature-generating capability of the protein coat and membrane tension that can predict whether the coat alone is sufficient to produce closed buds. Additionally, we show that an applied force, simulating a force from actin polymerization, can overcome the substantial energy difference that exists between ``open'' and ``closed'' buds at physiologically relevant membrane tensions. These findings demonstrate that membrane tension is an important determinant of membrane morphology in CME, and these results are general to other membrane budding processes.

\subsubsection*{Author Contributions}

J.E.H., G.O., and P.R. designed research. J.E.H. performed research. J.E.H., D.G.D., and P.R. analyzed data. J.E.H., G.O., D.G.D., and P.R. wrote the paper.

\newpage

\section*{Introduction}

The plasma membrane of animal cells is under tension as a result of in-plane stresses in the bilayer and connections between the membrane and the underlying actomyosin cortex \cite{Hochmuth1996,DizMunoz2013}. In recent years, it has become increasingly clear that this membrane tension plays an important role in a variety of cellular processes, from cell motility \cite{Houk2012} to controlling the balance of exocytosis and endocytosis \cite{Dai1997,Gauthier2011}. Of particular relevance to membrane-remodeling processes like endocytosis is the fact that membrane tension opposes deformations to the membrane by curvature-generating proteins \cite{Shi2015}. In fact, despite estimates that membrane tension should have negligible impact on the energetics of membrane budding \cite{Stachowiak2013}, recent experiments have shown that elevated membrane tension can inhibit clathrin-mediated endocytosis \cite{Boulant2011,Saleem2015}.

\begin{figure}

\includegraphics[width= \linewidth]{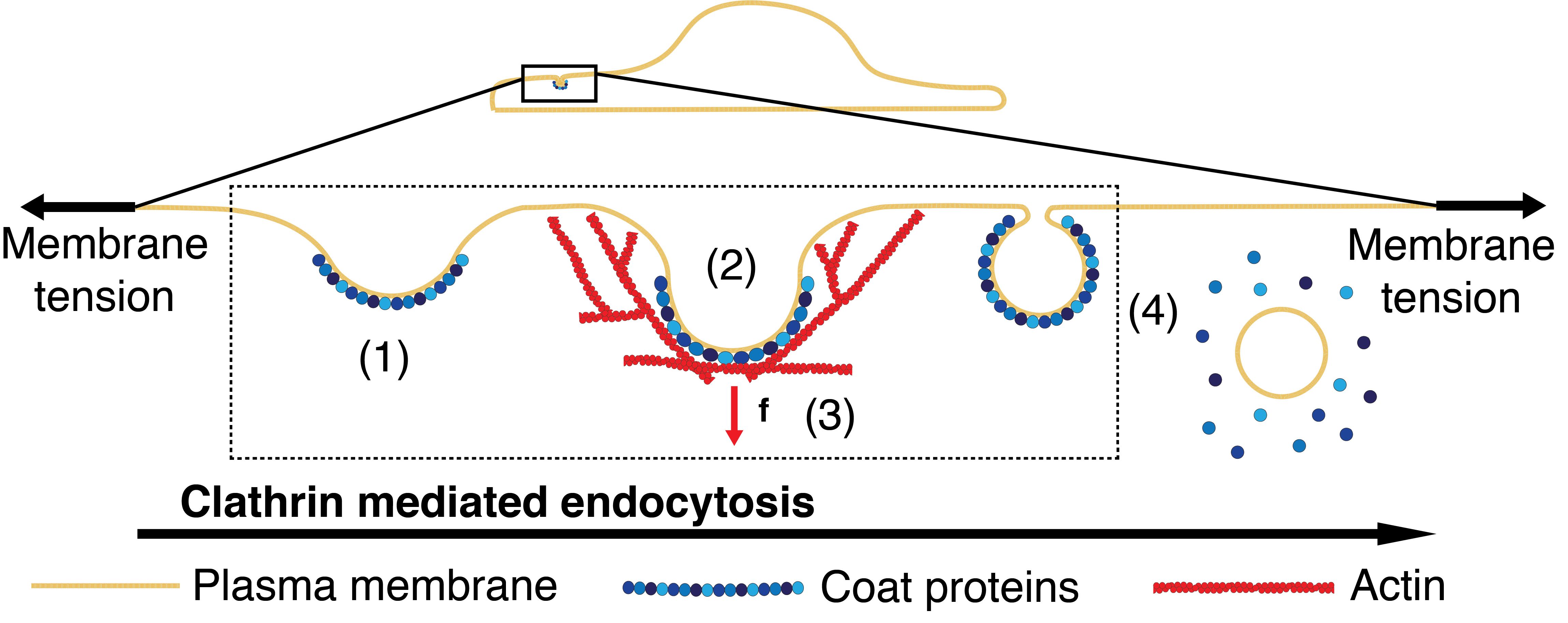}

\caption{Schematic depiction of the key mechanical steps in clathrin-mediated endocytosis. (1) A multi-component \textcolor{myBlue}{protein coat} forms on the \textcolor{myYellow}{plasma membrane} and induces the membrane to bend inwards, forming a shallow pit. (2) As the coat grows, the membrane becomes deeply invaginated to form an open, U-shaped pit before constricting to form a closed, $\Omega$-shaped bud. The plasma membrane is under tension in physiological conditions, which opposes the budding process. (3) It is believed that \textcolor{BrickRed}{actin} polymerization provides a \textcolor{red}{force}, $\fb$, to facilitate these morphological changes, particularly at high membrane tensions \cite{Boulant2011}. (4) The bud subsequently undergoes scission to form an internalized vesicle and the coat is recycled. Our study is concerned with the impact of membrane tension on the morphological changes effected by the coat and actin polymerization, as indicated by the dashed box.}
\label{fig:schematic}

\end{figure}

Clathrin-mediated endocytosis (CME) is an essential cellular process in eukaryotes that is important for the intake of nutrients, signaling, and recycling the lipid and protein components of the plasma membrane \cite{Doherty2009}. During CME, over 60 different protein species act in a coordinated manner to invaginate a patch of the plasma membrane into a bud that subsequently undergoes scission, forming an internalized vesicle \cite{McMahon2011, Kirchhausen2014}. CME is a classical example of a mechanochemical process, where a feedback between the biochemistry of the protein machinery and the mechanics of the plasma membrane and the actin cytoskeleton control endocytic patch topology and morphology \cite{Liu2009, Liu2010}. Figure \ref{fig:schematic} outlines the key mechanical steps of this process.

%\subsection*{Protein machinery of CME}

A critical step in CME is the assembly of a multicomponent protein coat that clusters cargo and bends the membrane into a budded morphology. Clathrin assembles into a lattice-like cage on the membrane with the assistance of adaptor proteins that directly bind lipids \cite{Kirchhausen2014, McMahon2011}. This assembly is generally thought to act as a scaffold that imposes its spontaneous curvature on the underlying membrane \cite{Dannhauser2012}. %However, it has been suggested that, because \revise{clathrin binds to the flexible region of most adaptor proteins, the potential force generated by polymerization would be inefficiently transmitted to deform the plasma membrane.} \cite{McMahon2011}.  
Recent work suggests that other components of the coat can also contribute to membrane bending via scaffolding by F-BAR domains, amphipathic helix insertion into the bilayer, and adaptor protein crowding \cite{Ford2002, Kirchhausen2014, Stachowiak2012, Stachowiak2013, Busch2015}.
The contributions from each of these membrane bending mechanisms can be combined into a single measure of the curvature generating capability of the coat, or spontaneous curvature, with an effective strength that depends on its composition, density and area coverage \cite{Helfrich1973, Lipowsky2013}.

%\subsection*{Mechanical forces in CME}
While coat proteins promote membrane budding, this process is opposed by the bending rigidity of the plasma membrane, membrane tension, turgor pressure, and cargo crowding \cite{Stachowiak2013, Boulant2011, Basu2013, Busch2015}. Crowding of cargo molecules on the outer leaflet of the plasma membrane opposes invagination of the membrane \cite{Stachowiak2013, Busch2015, Miller2015}, though we can think of this effect as simply a negative contribution to the spontaneous curvature of the coat. In yeast, turgor pressure is believed to be the primary opposing force to endocytosis due to their especially high internal pressure \cite{Basu2013, Dmitrieff2015}. However, in mammalian cells, the turgor pressure is several orders of magnitude lower than that in yeast \cite{Dai1999}, and therefore contributes comparatively little as an opposing force to CME in these cells. Consequently, membrane bending resistance is generally thought to be the primary opposing force to membrane budding \cite{Stachowiak2013}, though recent work has shown that membrane tension may also be a significant factor in this process \cite{Boulant2011,Saleem2015}.

%Bending rigidity is generally thought to be the primary opposing force to membrane deformations. The rigidity of the plasma membrane is a material property of the lipid bilayer describing its resistance to bending, and is determined by its composition \cite{Stachowiak2013, others}. 

\emph{In vivo}, elevated tension in combination with actin inhibitors causes clathrin-coated pits (CCPs) to exhibit longer lifetimes and increased the number of long-lived, presumably stalled, pits \cite{Boulant2011}. Under these conditions, open, U-shaped pits were found to be enriched as compared to closed, $\Omega$-shaped pits when visualized by electron microscopy \cite{Boulant2011, Yarar2005}. Similar observations have been made \emph{in vitro} where purified coat proteins were able to substantially deform synthetic lipid vesicles under low tension but were stalled at shallow, U-shaped pits at a higher tension \cite{Saleem2015}. These studies demonstrated that elevated membrane tension can block the progression of CME and that actin polymerization seems to be necessary to overcome this opposition. However, these studies did not fully address the magnitude at which membrane tension becomes important relative to the curvature-generating capability of the coat nor the magnitude of applied force necessary to overcome elevated tension.

There are many challenges associated with elucidating the effect of membrane tension on CME using experimental techniques. The diffraction-limited size of CCPs ($\sim100 \,\mathrm{nm}$) makes it currently impossible to directly image the morphology of the membrane in a live cell. The regularity of yeast CME has allowed for the visualization of time-resolved membrane shapes in this organism via correlative fluorescence and electron microscopy \cite{Kukulski2012, Picco2015}. However, the wide distribution of CCP lifetimes in mammalian cells \cite{Grassart2014, Aguet2013} makes this approach difficult in these cells. Additionally, current techniques are only capable of measuring global tension \cite{Dai1995, Dai1998, DizMunoz2013}. Due to these limitations, it is impossible to determine how the local membrane tension at a given CCP impacts the progression of membrane deformation.

On the other hand, mathematical modeling has provided insight into various aspects of membrane deformation in CME. For example, Liu et al. showed that a line tension at a lipid phase boundary could drive scission in yeast \cite{Liu2006, Liu2009}, while Walani et al. showed that scission could be achieved via snapthrough transition at high membrane tension \cite{Walani2015}. These studies and others \cite{Dmitrieff2015, Carlsson2014, Zhang2015} have demonstrated the utility of membrane modeling approaches for studying CME, though none have systematically explored the effect of varying membrane tension on the morphological progression of membrane budding.

Since an endocytic patch is a small region ($\sim 100 \,\mathrm{nm}$ \cite{Collins2011, Avinoam2015}) that is connected to the larger cell membrane, we reasoned that plasma membrane tension is likely to be key determinant of the bud morphology in CME. In this study, we develop a modeling framework to investigate the impact of membrane tension on budding processes like CME. This model accommodates heterogeneous membrane composition through the spontaneous curvature term. Using this model we show that membrane tension, in conjunction with the coat spontaneous curvature and coat coverage area, determines the morphology of the CCP. We also identify the circumstances in which CCPs traverse a \emph{snapthrough instability} to transition from U-shaped buds to $\Omega$-shaped buds. Finally, we show that an externally applied force can mediate this transition.

%%%%%%%%%%%%%%%%%%%%%%%%%%%%%%%%%%%%%%%%%%%

\section*{Model development}

%%%%%%%%%%%%%%%%%%%%%%%%%%%%%%%%%%%%%%%%%%%

%\subsection*{Assumptions}
We model the lipid bilayer as a thin elastic shell. The bending energy of the membrane is modeled using the Helfrich-Canham energy, which is valid for radii of curvatures much larger than the thickness of the bilayer \cite{Helfrich1973}. Since typical endocytic patch radii of curvatures are $\approx 50 \,\mathrm{nm}$ \cite{Collins2011, Avinoam2015}, application of this model provides a valid representation of the shapes of the membrane. Further, we assume that the membrane is at mechanical equilibrium at all times. Because clathrin-mediated endocytosis occurs over a timescale of tens of seconds \cite{Boulant2011,Taylor2011, Aguet2013, Grassart2014}, the membrane has sufficient time to attain mechanical equilibrium at each stage \cite{Liu2009, Dmitrieff2015}. We also assume that the membrane is incompressible because the energetic cost of stretching the membrane is high \cite{Rawicz2000}. This incompressibility constraint is implemented using a Lagrange multiplier (see Section \ref{sec:extended_model} SOM for details). Finally, for simplicity in the numerical simulations, we assume that the endocytic pit is rotationally symmetric (Figure \ref{fig:geometry}).

Since one of the key features of CME is coat-protein association with the plasma membrane, we model the strength of curvature induced by the coat proteins with a spontaneous curvature term ($C$). The spontaneous curvature represents an asymmetry across the membrane that favors bending in one direction over the other with a magnitude equal to the inverse of the preferred radius of curvature. Classically, this was used to represent differences in the lipid composition (i.e. head group size and tail length/number) of the two monolayers that would cause the membrane to bend \cite{Helfrich1973}. In our case, the spontaneous curvature represents the preferred curvature of the coat proteins bound to the cytosolic face of the membrane, consistent with its usage in other studies \cite{Rangamani2014,Walani2015,Dmitrieff2015,Shi2015,Bassereau2014}.

Our model reflects the fact that the clathrin coat covers a finite area and that this region should have different physical properties (e.g. spontaneous curvature, bending rigidity) than the surrounding membrane. Heterogeneity in the spontaneous curvature and bending rigidity is accommodated by using a local rather than global area incompressibility constraint \cite{Agrawal2009,Steigmann2003}. This allows us to simulate a clathrin coat by tuning the size, spontaneous curvature, and rigidity of the ``coated'' region relative to the bare membrane.

%%%%%%%%%%%%%%%%%%%%%%%%%%%%%%%%%%%%%%%%%%%%

We use a modified version of the Helfrich energy that includes a spatially-varying spontaneous curvature, $C \left(\theta^{\alpha} \right)$ \cite{Rangamani2014,Walani2015,Agrawal2009},
\begin{align} \label{eq:Helfrich}
W = k \left( H - C(\theta^{\alpha}) \right)^2 + \bar{k} K,
\end{align}

\noindent where $W$ is the energy per unit area, $k$ is the bending modulus, $\bar{k}$ is the Gaussian bending modulus, $H$ is the local mean curvature, and $K$ is the local Gaussian curvature. $\theta^{\alpha}$ denotes the surface coordinates where $\alpha \in $ \{1,2\}. Note that this energy function differs from the standard Helfrich energy by a factor of $2$, with the net effect being that our value for the bending modulus, $k$, is twice that of the standard bending modulus typically  encountered in the literature.  The resulting ``shape equation'' for this energy functional is 
\begin{equation} 
\underbrace{k \Delta \left(H - C \right) + 2 k \left( H - C \right) \left(2 H^2 - K \right) - 2 k H \left( H - C \right)^2}_{\text{Elastic Effects}} = \underbrace{p + 2 \lambda H}_{\text{Capillary effects}} + \underbrace{\fb \cdot \nb}_{\text{Force due to actin}},
\label{eq:shape}
\end{equation}

\noindent where $\Delta$ is the surface Laplacian, $p$ is the pressure difference across the membrane, $\lambda$ is interpreted to be the membrane tension, $\mathbf{f}$ is a force per unit area applied to the membrane surface, and $\mathbf{n}$ is the unit normal to the surface \cite{Agrawal2009, Walani2015}. In this model, $\fb$ represents the applied force exerted by the actin cytoskeleton; this force need not necessarily be normal to the membrane.

A consequence of heterogenous protein-induced spontaneous curvature and externally applied force is that $\lambda$ is not homogeneous in the membrane \cite{Agrawal2009,Rangamani2014}. The spatial variation in $\lambda$ is accounted for as
\begin{equation} 
\underbrace{\lambda_{, \alpha}}_{\text{Gradient of surface pressure}} = \underbrace{- 2 k \left( H - C \right) \frac{\partial{C}}{\partial{x^{\alpha}}}}_{\text{protein-induced variation}} - \underbrace{\fb\cdot\ab_{\alpha}}_{\text{force induced variation}},
\label{eq:lambda}
\end{equation}

\noindent where $\left(\cdot\right)_{, \alpha}$ is the partial derivative with respect to the coordinate $\alpha$ and $\ab_{\alpha}$ is the unit tangent in the $\alpha$ direction. $\lambda$ can be interpreted as the surface tension \cite{Rangamani2014,Steigmann1999}, and in our case is also affected by the tangential components ($\ab_{\alpha}$) of the force due to the actin cytoskeleton. The complete derivation of the boundary value problem and simulation details are given for axisymmetric coordinates in the SOM. 

\section*{Results}

%%%%%%%%%%%%%%%%%%%%%%%%%%%%%%%%%%%%%%%%%%%

\subsection*{Membrane tension inhibits bud formation by curvature-generating coats} \label{ss:tension}

\begin{figure}[t]

\includegraphics[width= \textwidth]{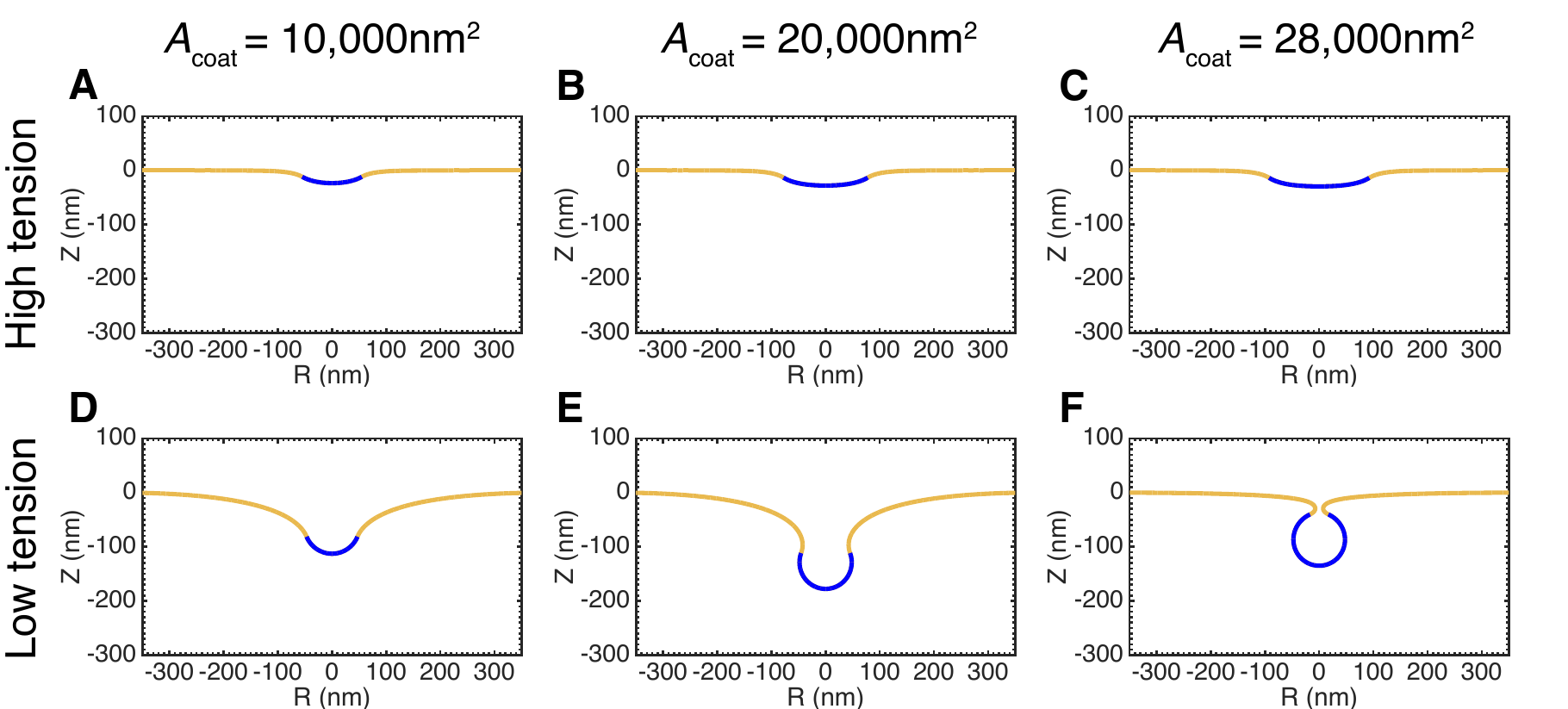} 

\caption{Profile views of membrane morphologies generated by simulations where the area of a \textcolor{blue}{curvature-generating coat} progressively increases, covering more of the \textcolor{myYellow}{bare membrane}. Simulations were performed for both high and low  values of membrane tension. The curvature-generating capability, or spontaneous curvature, of the coat is set at $C_0 = 0.02\, \mathrm{nm}^{-1}$, corresponding to a preferred radius of curvature of $50 \,\mathrm{nm}$. \textbf{(C)} High membrane tension, $\lambda_0 = 0.2 \, \textrm{pN}/\textrm{nm}$. The membrane remains nearly flat as the area of the coat increases. \textbf{(D-F)} Low membrane tension, $\lambda_0 = 0.002 \, \textrm{pN}/\textrm{nm}$. Addition of coat produces a smooth evolution from a flat membrane to a closed bud.}

\label{fig:tensionComp}

\end{figure}

In order to understand how membrane tension affects the morphology of a coated membrane, we performed simulations in which a curvature-generating coat ``grows'' from the center of an initially flat patch of membrane. For simplicity, the spontaneous curvature was set to be constant, $C_0 = 0.02 \,\mathrm{nm}^2$, in the coated region with a sharp transition at the boundary between the coated and bare membrane (implemented via hyperbolic tangent functions, Figure \ref{fig:tanh}). The membrane tension was varied by setting the value of $\lambda$ at the boundary of the membrane patch, which corresponds to the tension in the surrounding membrane reservoir. Figure \ref{fig:tensionComp} shows the morphology of the membrane under low and high tension conditions for these ``coat-growing'' simulations, in the absence of any force from actin assembly (i.e $\fb=0$).

%%%%%%%%%%%%%%%%%%%%%%%%%%%%%%%%%%%%%%%%%%%

%\subsubsection*{Increasing coat area}

High membrane tension ($0.2 \,\mathrm{pN/nm}$) inhibits deformation of the membrane by the protein coat (Figure \ref{fig:tensionComp}C). Even as the area of the coated region ($A_{\mathrm{coat}}$) increases, the membrane remains nearly flat. In fact, the size of the coated region can grow arbitrarily large without any substantial deformation (Figure \ref{fig:highTensionBigCoat}). The spontaneous curvature of the coat is simply unable to overcome the additional resistance provided by the high membrane tension. 

In contrast, at low membrane tension ($0.002 \,\mathrm{pN/nm}$), the protein coat is able to substantially deform the membrane from its initial, flat morphology to a closed bud(Figure \ref{fig:tensionComp}D-F). Increasing the coat area causes a smooth evolution from a shallow to deep U-shape to a closed, $\Omega$-shaped bud. We stopped the simulations when the membrane is within $5$ nm of touching at the neck, at which point bilayer fusion resulting in vesicle scission is predicted to occur spontaneously \cite{Liu2006}. These morphological changes are similar to those observed in clathrin-mediated endocytosis \cite{Avinoam2015} and do not depend on the size of the membrane patch (Figure \ref{fig:alphaComp}).

%%%%%%%%%%%%%%%%%%%%%%%%%%%%%%%%%%%%%%%%%%%

%\subsubsection*{Increasing coat spontaneous curvature}

\begin{figure}[t]

\includegraphics[width= \textwidth]{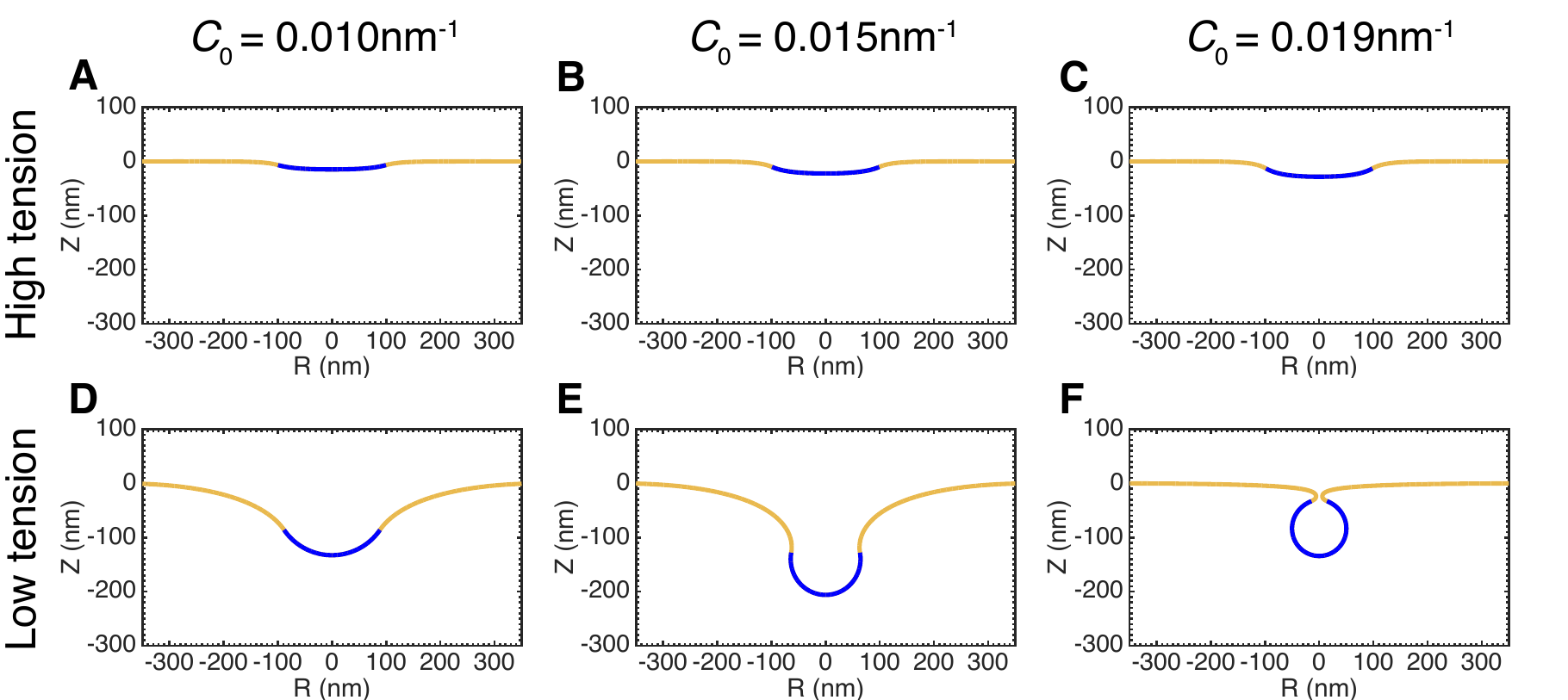}

\caption{Membrane profiles for simulations with a constant coat area in which the spontaneous curvature of the \textcolor{blue}{coat} progressively increases. The area of the coat is $31,416 \,\textrm{nm}^2$, the approximate surface area of a clathrin-coated vesicle \cite{Avinoam2015}. \textbf{(C)} High membrane tension, $\lambda_0 = 0.2 \, \textrm{pN}/\textrm{nm}$. The membrane remains nearly flat with increasing spontaneous curvature. \textbf{(D-F)} Low membrane tension, $\lambda_0 = 0.002 \, \textrm{pN}/\textrm{nm}$. Increasing the spontaneous curvature of the coat produces a smooth evolution from a flat membrane to a closed bud.}

\label{fig:c0tensionComp}

\end{figure}

Can increasing the spontaneous curvature of the coat overcome tension-mediated resistance to deformation? To answer this question, we performed simulations in which the spontaneous curvature of the coat increases while the area covered by the coat remains constant at approximately the surface area of a typical clathrin coated vesicle \cite{Avinoam2015}. As before, high membrane tension (Figure \ref{fig:c0tensionComp}C) prevents deformation of the membrane by the coat. Even increasing the spontaneous curvature to a value of $0.04$ nm$^{-1}$, corresponding to a preferred radius of curvature of $25$ nm and twice the value used in the simulations in Figure \ref{fig:tensionComp}, does not produce a closed bud (Figure \ref{fig:endoCurvHighTension}). In the case of low membrane tension (Figure \ref{fig:c0tensionComp}D-F), a progressive increase in the coat spontaneous curvature causes a smooth evolution from a shallow to deep U-shape to a closed, $\Omega$-shaped bud. 

The similarity between the membrane morphologies in Figures \ref{fig:tensionComp} and \ref{fig:c0tensionComp} suggests that the interplay between spontaneous curvature, coat area and membrane tension is more important than the exact timing of coat protein arrival relative to the bending of the membrane, particularly in the absence of other membrane bending mechanisms. In the results that follow, we performed simulations with increasing coat area, though the results are equally applicable to the case of increasing coat spontaneous curvature. 

%%%%%%%%%%%%%%%%%%%%%%%%%%%%%%%%%%%%%%%%%%%

\subsection*{Transition from U- to $\Omega$-shaped buds occurs via  instability at intermediate membrane tensions} \label{ss:instability}

\begin{figure}[t]

\includegraphics[width=\textwidth]{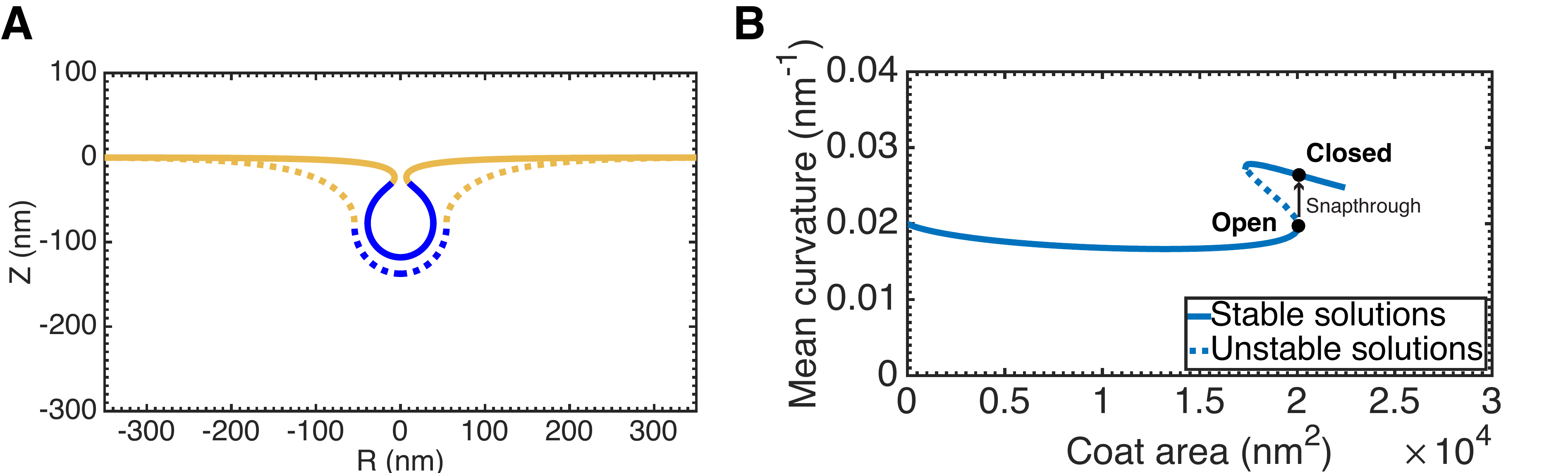}

\caption{A snapthrough instability exists for increasing coat area at intermediate, physiologically relevant \cite{Sens2015}, membrane tensions, $\lambda_0 = 0.02 \, \textrm{pN}/\textrm{nm}$. \textbf{(A)} Membrane profiles showing bud morphology before (dashed line, $A_{\mathrm{coat}} = 20{,}065 \, \mathrm{nm}^2$) and after (solid line, $A_{\mathrm{coat}} = 20{,}105 \, \mathrm{nm}^2$ ) addition of a small amount of area to the coat. \textbf{(B)} Mean curvature at the tip of the bud as a function of the coat area.  There are two stable branches of solutions of the equilibrium membrane shape equations. The lower branch consists of open, U-shaped buds while the upper branch consists of closed, $\Omega$-shaped buds. The dashed portion of the curve indicates ``unstable'' solutions that are not accessible by simply increasing and decreasing the area of the coat. The marked positions on the curve denote the membrane profiles shown in (A). The transition between these two shapes is a snapthrough, in which the bud ``snaps'' closed upon a small addition to area of the coat.}

\label{fig:snapthrough}

\end{figure}

Experimentally measured membrane tensions in live mammalian cells typically fall between the high and low tension regimes presented in Figures \ref{fig:tensionComp} and \ref{fig:c0tensionComp} \cite{Sens2015}.  Figure \ref{fig:snapthrough} shows the result of a coat-growing simulation with membrane tension at an intermediate value of $0.02$ pN/nm. As in the low membrane tension case, increasing the area of the coat causes substantial deformation of the membrane. However, there is no longer a smooth transition from an open bud to a closed bud. Figure \ref{fig:snapthrough}A shows a bud just before (dashed line) and after (solid line) a small amount of area is added to the coat. Evidently, this small change causes the bud to ``snap'' closed to an $\Omega$-shaped morphology. This situation is known as a  \emph{snapthrough instability}, and similar instabilities have been observed in other recent membrane modeling studies \cite{Walani2015, Dmitrieff2015}. We emphasize that these are two equilibrium shapes of the membrane, and the exact transition between these states (i.e. intermediate unstable shapes and timescale) is not modeled here.

To better visualize why this sharp transition should occur, Figure \ref{fig:snapthrough}B shows the mean curvature at the tip of the bud as a function of the area of the coat. In comparison to the high and low membrane tension cases (Figure \ref{fig:HvsA0}), there are two branches of stable solutions. The lower and upper branches represent ``open'' and ``closed'' morphologies of the bud, respectively. The marked solutions show the location of the two morphologies depicted in Figure \ref{fig:snapthrough}A. The open bud in Figure \ref{fig:snapthrough}A is at the end of the open bud solution branch, so any addition of area to the coat necessitates that the membrane adopt a closed morphology. A similar snapthrough, albeit with a less severe morphological transition, occurs when the bending rigidity of the coated region is increased \cite{Jin2006} relative to the bare membrane (Figure \ref{fig:stiffCoat}).

\begin{figure}[t]

\includegraphics[width=\textwidth]{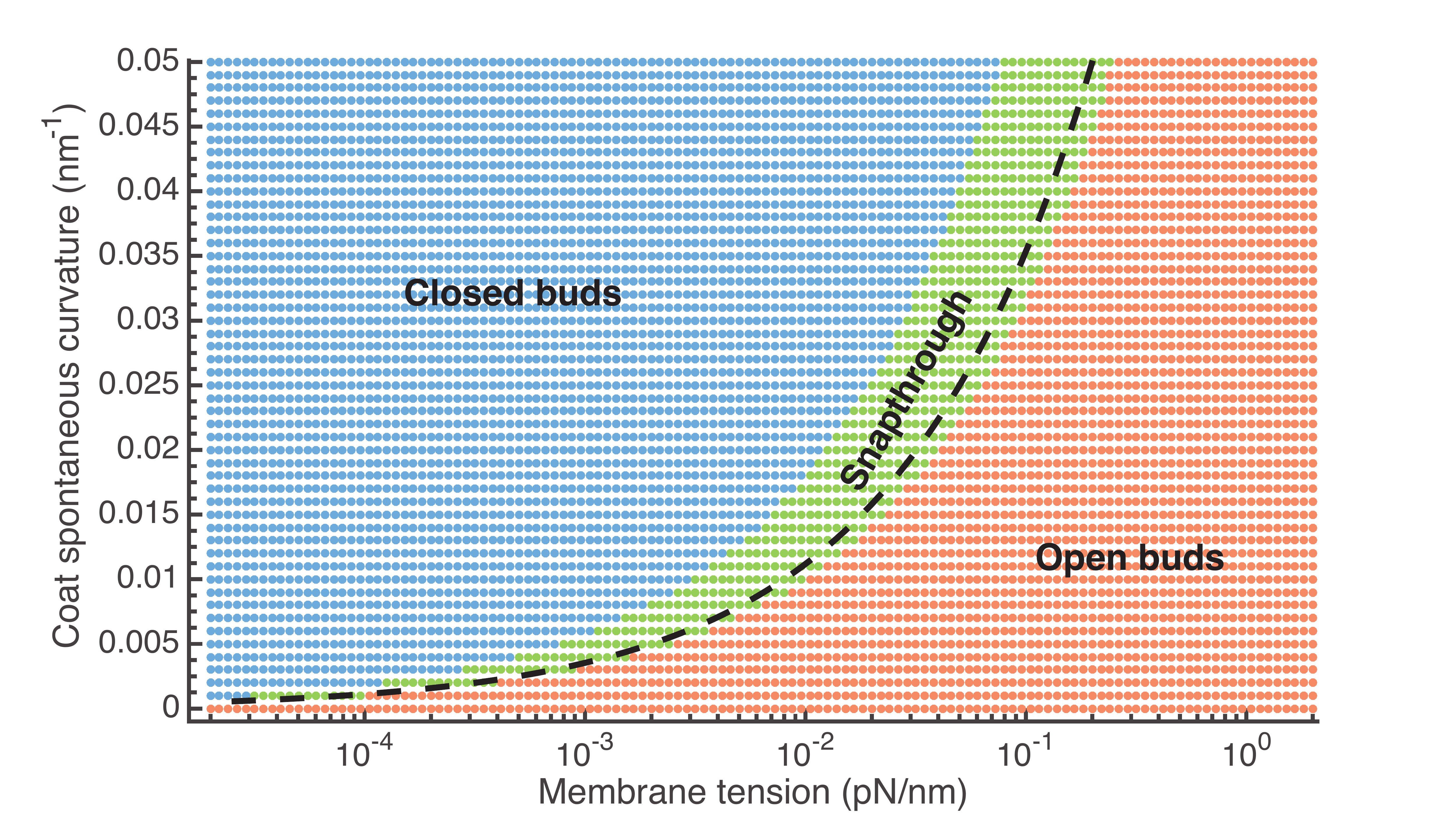}

\caption{Coat spontaneous curvature ($C_0$) vs. membrane tension ($\lambda_0$) phase diagram. The shape evolution of the budding process depends on both the membrane tension and coat spontaneous curvature. Each dot represents a coat ``growing'' simulation performed with the specified values for edge membrane tension and coat spontaneous curvature. The dots are colored according to the final shape of the membrane: \textcolor{myBlue}{Blue} denotes closed, $\Omega$-buds, \textcolor{myRed}{Red} denotes open, U-shaped pits, and \textcolor{myGreen}{Green} are situations in which closed buds are obtained via a snapthrough transition. The snapthrough solutions cluster about the dotted line, $\mathrm{Ves} = 1$, which separates the ``high'' and ``low'' membrane tension regimes (see main text).
}

\label{fig:phase}

\end{figure}

Over what ranges of tension and spontaneous curvature does this snapthrough transition occur? In order to understand the nature of the transition between low and high membrane tension regimes, we performed simulations over several orders of magnitude of the membrane tension ($10^{-4}$ to $1$ pN/nm), encompassing the entire range of measured physiological tensions \cite{Sens2015} (Figure \ref{fig:cusp}). We then performed these simulations over a range of spontaneous curvatures of the coat ($0$ to $0.05$ $\mathrm{nm}^{-1}$) corresponding to preferred radii of curvature from $20$ nm and up. Based on the results, we constructed a phase diagram summarizing the observed morphologies (Figure \ref{fig:phase}). Each dot in the diagram represents one simulation at the corresponding values of membrane tension and coat spontaneous curvature. The \textcolor{myBlue}{blue} region denotes a smooth evolution to a closed bud, the \textcolor{myRed}{red} region represents a failure to form a closed bud, and \textcolor{myGreen}{green} region indicates a snapthrough transition from an open to a closed bud. This phase diagram clearly shows that the distinction between ``low'' and ``high'' membrane tension conditions depends on the magnitude of the spontaneous curvature of the coat.

These results can be understood by comparing the spontaneous curvature of the coat to the membrane tension and bending rigidity via the dimensionless quantity, $\mathrm{Ves} = \frac{C_0}{2} \sqrt{\frac{\kappa}{\lambda}}$, hereafter termed the \emph{vesiculation number}. The dashed line in Figure \ref{fig:phase} corresponds to $\mathrm{Ves} = 1$, which bisects the low ($\mathrm{Ves} > 1$) and high tension ($\mathrm{Ves} < 1$) results. The snapthrough results cluster about this line, marking the transition region between the high and low tension cases. Importantly, this demonstrates that the preferred radius of curvature of the coat, $1/C_0$, must be smaller than the ``natural'' length scale of the membrane,  $\frac{1}{2}\sqrt{\kappa/\lambda}$ \cite{Dmitrieff2015}, for the coat to be capable of producing a closed bud in the absence of other mechanisms of curvature generation. 

\subsection*{A large energy barrier accompanies the instability} \label{ss:energy}

\begin{figure}[t]

\centering

\includegraphics[width=0.75\textwidth]{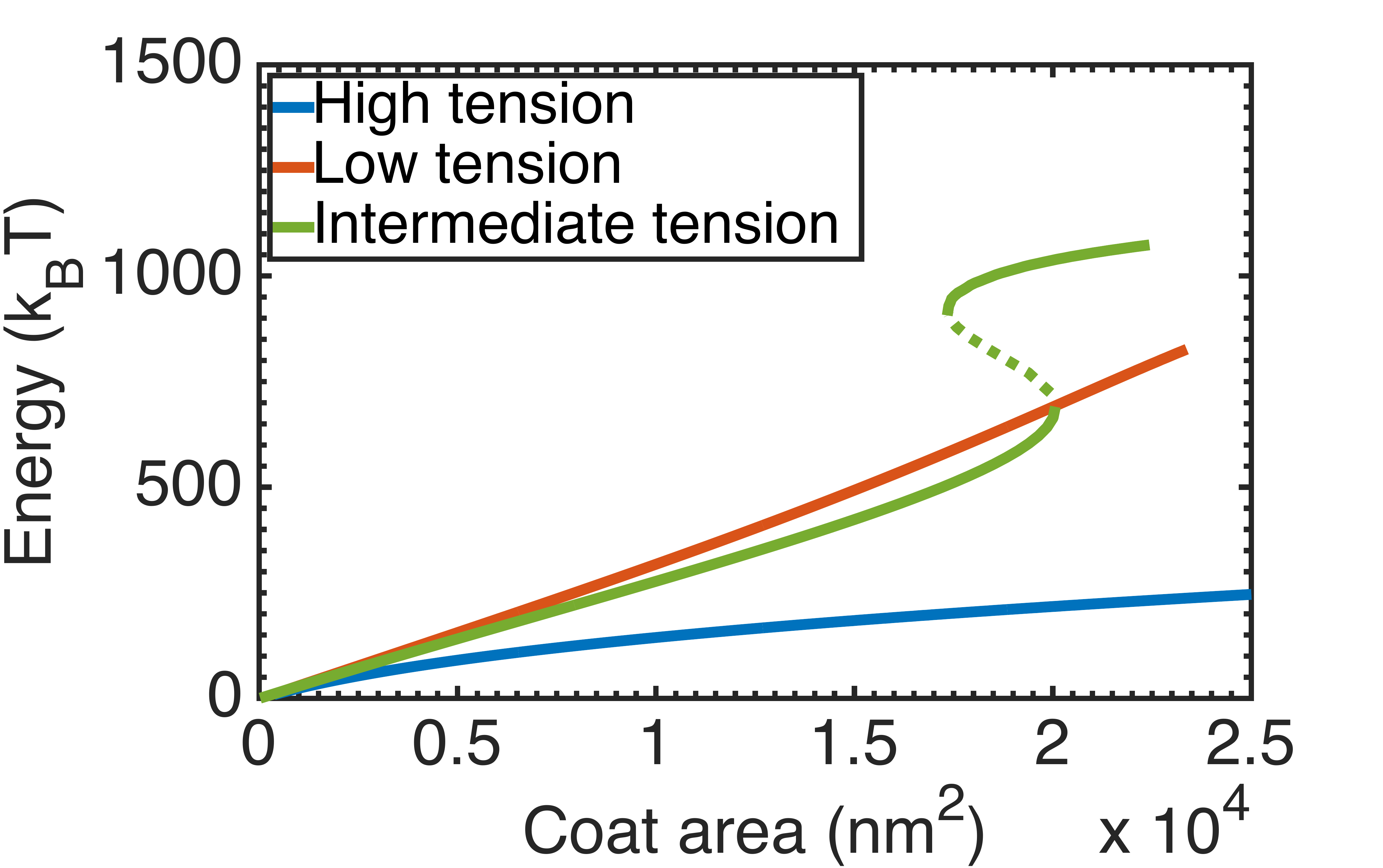}

\caption{An energy barrier opposes coat closing at the instability. The energy input necessary to deform the membrane is plotted as a function of the area of the coat for three values of membrane tension. \textcolor{myBlue}{High membrane tension}, $\lambda_0 = 0.2 \, \textrm{pN}/\textrm{nm}$. \textcolor{myRed}{Low membrane tension}, $\lambda_0 = 0.002 \, \textrm{pN}/\textrm{nm}$. \textcolor{myGreen}{Intermediate membrane tension}, $\lambda_0 = 0.02 \, \textrm{pN}/\textrm{nm}$. At the intermediate membrane tension, there is an energy difference of over $100 \,\mathrm{k_B T}$ between the U- and $\Omega$-shaped buds. This energy barrier could prevent the transition from U- to $\Omega$-shaped buds if the polymerization energy of the coat is insufficient to overcome the barrier.}

\label{fig:energy}

\end{figure}

What is the energy cost that is associated with membrane deformation? For the budding process to proceed from an open, shallow invagination to a closed bud, the free energy gain from the assembly of the coat (\emph{i.e.} the binding energy of the protein-protein and protein-lipid interactions) must exceed the energy necessary to deform the membrane \cite{Zimmerberg2006, Saleem2015}. To examine this requirement, we calculated the energy needed to bend the membrane at the low, high, and intermediate membrane tensions as a function of the coat area (Figure \ref{fig:energy}). Details of the calculation can be found in Section \ref{s:energy} of the SOM. Since the energy is proportional to the square of the mean curvature (Eq. \ref{eq:Helfrich}), it is not surprising that the highly curved, budded morphologies in the low membrane tension case cost more energy to produce than the flat morphologies at high tension. Indeed, the main contribution to the energy in the low tension case is the bending ridigity (Figure \ref{fig:eVsCarea}C), whereas the energy cost from tension and bending rigidity are comparable at high tension (Figure \ref{fig:eVsCarea}A).

In the case of intermediate membrane tension, we again see the two stable solution branches corresponding to the open and closed bud morphologies. Notably, the major opposing force is bending rigidity of the membrane and resistance from membrane tension contributes only a minor amount to the overall energy (Figure \ref{fig:eVsCarea}B), consistent with previous estimates \cite{Stachowiak2013}. Though the energy for the open buds in the intermediate case is similar to that in the low tension case, there is a large ($>100 \,\mathrm{k_B T} \approx 250 \,\mathrm{kJ/mol} $) energy barrier between the open and closed bud morphologies. Thus, \emph{if} the polymerization energy of the coat is insufficient to reach the upper solution branch, then we should expect to observe ``stalled'' buds in an open, U-shaped configuration at intermediate membrane tensions in the absence of other mechanisms of curvature generation. It should be noted that the exact magnitude of the energy necessary to deform the membrane, along with the energy barrier at this intermediate membrane tension, depends on the bending rigidity, $k_0$, of the membrane (see Equation \ref{eq:Ebend}). 

\begin{figure}[t]

\includegraphics[width=\textwidth]{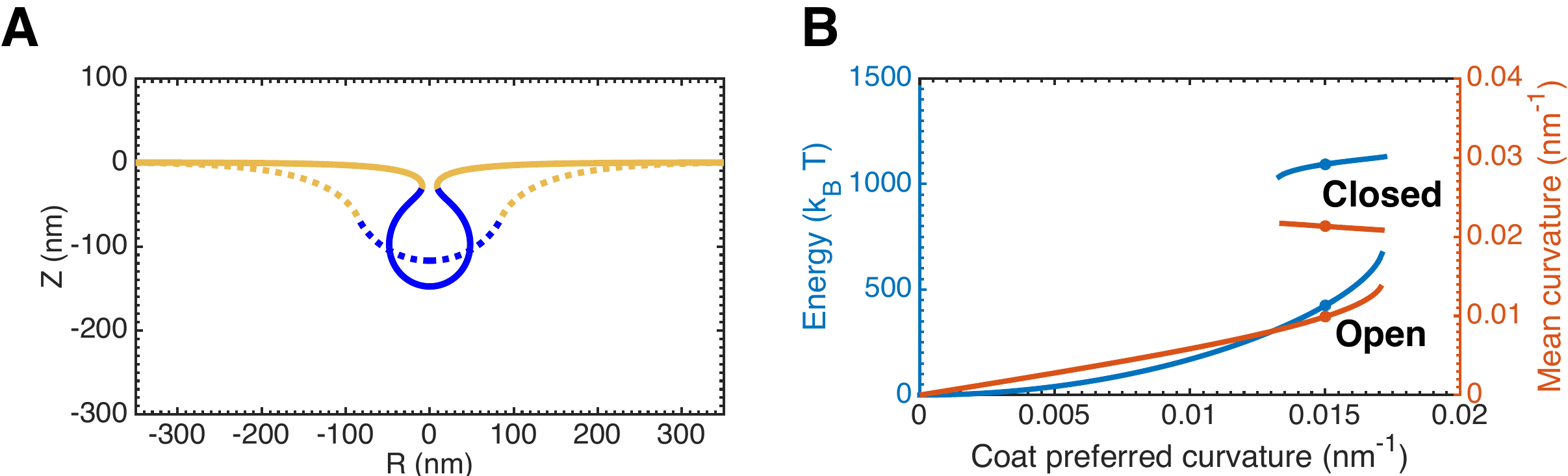}

\caption{The instability and energy barrier are present at intermediate tensions for increasing spontaneous curvature $C_0$ , $\lambda_0 = 0.02 \, \textrm{pN}/\textrm{nm}$, $A_{\mathrm{coat}} = 31,416 \,\textrm{nm}^2$. \textbf{(A)} Membrane profiles showing U-shaped (dashed line) and $\Omega$-shaped (solid line) solutions at an identical coat spontaneous curvature, $C_0 = 0.015 \mathrm{nm}^2$. \textbf{(B)} \textcolor{myBlue}{Energy necessary to deform the membrane} and \textcolor{myRed}{mean curvature at the tip of the bud} as a function of the spontaneous curvature of the coat. The two stable branches are open (lower) and closed (upper) buds. There is a substantial energy barrier ($>100 \,\mathrm{k_B T}$) between the U- and $\Omega$-shaped buds.}

\label{fig:endoCurv}

\end{figure}

We observed a similar result  when the spontaneous curvature was varied for a constant coat area. Figure \ref{fig:endoCurv}A shows representative open (dashed line) and closed (solid line) membrane morphologies at identical spontaneous curvatures. Figure \ref{fig:endoCurv}B plots the \textcolor{myBlue}{energy necessary to deform the membrane} and \textcolor{myRed}{mean curvature of the bud tip} as a function of the spontaneous curvature of the coat, and the marked points denote the solutions shown in Figure \ref{fig:endoCurv}A. There are again two branches of solutions to the equilibrium equations with the lower branch representing open buds and the upper branch closed buds. Substantially more energy ($>100 \,\mathrm{k_B T}$) is required to deform the membrane into closed buds as compared to the open buds at identical coat spontaneous curvatures.

%%%%%%%%%%%%%%%%%%%%%%%%%%%%%%%%%%%%%%%%%%%

%\subsection*{The instability exists over a range of membrane tensions and spontaneous curvatures} \label{ss:energy}

%%%%%%%%%%%%%%%%%%%%%%%%%%%%%%%%%%%%%%%%%%%

\subsection*{A pulling force can mediate the transition from a U- to $\Omega$-shaped bud} \label{ss:force}

\begin{figure}[tbp]

\includegraphics[width=\textwidth]{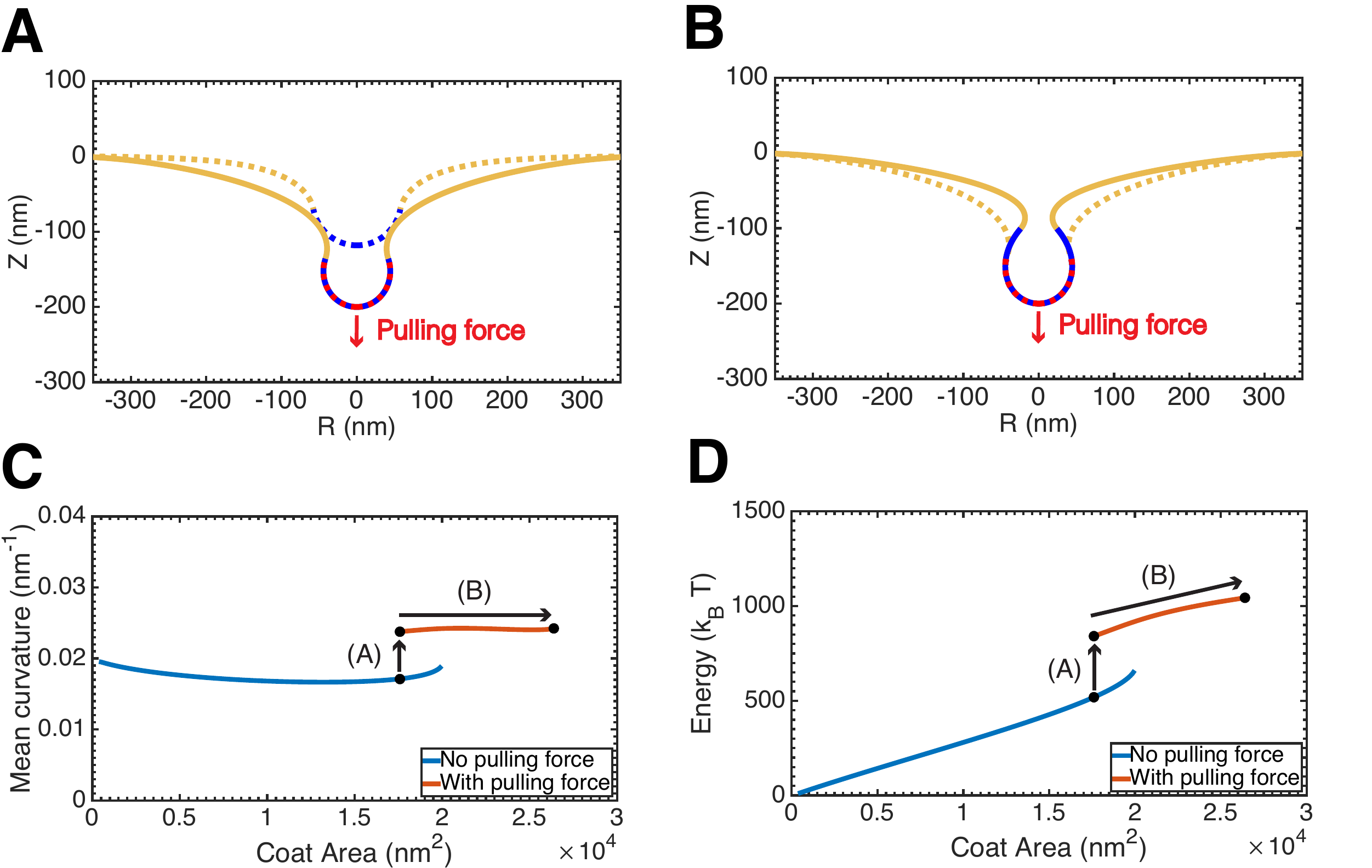}

\caption{A pulling force can mediate the transition from a U- to $\Omega$-shaped bud. \textbf{(A)} At a constant coat area, $A_{\mathrm{coat}}= 17,593 \,\mathrm{nm}^2$, a \textcolor{red}{pulling force} (red dash) localized to the coat drives the shape transition from a U-shaped (dashed line) to $\Omega$-shaped bud (solid line). \textbf{(B)} The bud can be closed further (solid line) by increasing the area of the coat while the \textcolor{red}{pulling force} holds the tip of the bud at $Z = -200 \, \mathrm{nm}$. \textbf{(C)} Mean curvature at the tip of the bud as function of the coat area. The pulling force bridges the two solution branches. \textbf{(D)} Energy to deform the membrane as a function of the coat area. The pulling force provide the energy input to reach the upper solution branch. The black arrows in (C) and (D) indicate the direction of the simulations in (A) and (B) with the marked points indicating the solutions corresponding to the shown membrane profiles.}

\label{fig:pull}

\end{figure}

What mechanisms of force generation enable the cell to overcome the energy barrier? Experiments have demonstrated that CME is severely affected by a combination of elevated tension and actin inhibition \cite{Boulant2011, Grassart2014}. To examine whether a force from actin polymerization is sufficient to induce a transition from open to closed bud morphologies, we modeled the force from actin polymerization as a ``pulling'' force on the bud as shown in Figure \ref{fig:schematic}. It should be noted that the ultrastructure of the actin cytoskeleton at CME sites in live cells, and hence the exact orientation of the applied force, is currently unknown. This merely represents a candidate orientation analogous to the force from actin polymerization in yeast CME \cite{Picco2015}.

We find that a pulling force of less than $10 \,\mathrm{pN}$ is sufficient to drive the membrane from an open to closed configuration (Figure \ref{fig:pull}). Figure \ref{fig:pull}A demonstrates that a downward pulling force on the coated region is sufficient to drive a transition from an open U-shaped bud to an $\Omega$-shaped bud. Then, while holding the bud at a prescribed depth, area can be added to the coat, further closing the bud (Figure \ref{fig:pull}B). The force required to pull the bud (Figure \ref{fig:pull}A) and then maintain its position (Figure \ref{fig:pull}B) is less than $10$ pN (Figure \ref{fig:FvsZpA0}), well within the capability of a polymerizing actin network \cite{Bieling2016}. 

Once again, there are two branches of solutions with increasing coat area, as seen in Figures \ref{fig:pull}C and \ref{fig:pull}D which plot the mean curvature of the tip of the bud and the energy necessary to deform the membrane, respectively. However, these solution branches are now fundamentally different as $\mathbf{f} = 0$ pN in the \textcolor{myBlue}{lower branch} (identical to Figure \ref{fig:snapthrough}), while $\mathbf{f} \approx 6$ pN in the \textcolor{myRed}{upper branch} (Figure \ref{fig:FvsZpA0}B). There is a substantial energy barrier between the two solution branches ($\approx 350  \,\mathrm{k_B T}$), though this could be overcome by an actin network composed of a few hundred actin monomers, assuming an energy efficiency of $\sim 5$\% \cite{Bieling2016}. This is quite reasonable given recent measurements in yeast,  in which the number of actin monomers was estimated to be on the order of thousands \cite{Picco2015, Sirotkin2010}. An effective ``pinching'' force, as suggested by the results of Collins et al. \cite{Collins2011}, is also sufficient to induce the transition from U- to $\Omega$-shaped buds (Figure \ref{fig:squeeze}). These results demonstrate that the actin cytoskeleton can overcome an energy barrier to forming a closed bud at physiological membrane tensions in mammalian cells.

%%%%%%%%%%%%%%%%%%%%%%%%%%%%%%%%%%%%%%%%%%%

\section*{Discussion}

Despite our extensive knowledge of the protein machinery and dynamics of CME \cite{Taylor2011}, we do not completely understand how membrane tension influences this process.  An important challenge in understanding the role of membrane tension is that it is difficult to measure and to interpret the effects of membrane tension. Traditionally, membrane tension measurements on live cells have been performed by measuring the force required to hold a tube pulled from the plasma membrane \cite{Dai1995,Sinha2011}. The tension obtained from this measurement is a combination of the in-plane tension in the membrane as well as the energy from membrane-to-cortex attachments (MCA) \cite{Dai1999, DizMunoz2013}. Whether the MCA impacts the effective tension felt at CCPs is not clear, but because the actin cortex is not homogeneous throughout the cell \cite{Honigmann2014} and can act as a diffusion barrier to lipids and transmembrane proteins \cite{Trimble2015, Fujiwara2016}, endocytic pits in different regions of cells might be subject to different effective membrane tensions. 

In this study, we investigated the role of membrane tension in governing the morphological landscape of CME and found that a combination of membrane tension and protein-induced spontaneous curvature governs the morphology of the endocytic pit (Figures \ref{fig:tensionComp}, \ref{fig:c0tensionComp}). Additionally, we found that at intermediate membrane tensions, the bud must go through a snapthrough to go from an open to closed configuration (Figure \ref{fig:snapthrough}). The full range of bud morphologies for different values of membrane tension and spontaneous curvature is shown in Figure \ref{fig:phase}. A key result from this work is that the vesiculation number can be used to identify the regime of tension and curvature-mediated effects that separates the closed and open bud morphologies. Finally, we found that a force modeling actin polymerization in CME can mediate the transition between open and closed buds at physiologically relevant membrane tensions. We believe these can results can explain the observations of a number of recent experimental studies.

There has been conflicting evidence as to whether actin is an essential component of the endocytic machinery in mammalian cells \cite{Boulant2011,Yarar2005,Grassart2014}. This is in contrast to the situation in yeast where actin assembly is absolutely required for productive CME, presumably to counteract the substantial turgor pressure in this organism \cite{Basu2013,Dmitrieff2015}. A hypothesis that follows from our work is that the membrane tension in a particular cell type should determine whether actin polymerization is required for productive CME.

This hypothesis is consistent with experimental observations. Boulant et al. found that treatment with jasplakinolide left CME unaffected in typical conditions, but the same treatment had a severe inhibitory effect in cells subjected to hypoosmotic shock or mechanical stretching \cite{Boulant2011}. One interpretation of this result is that physiological membrane tension was initially in the ``low'' tension regime, and the increase in membrane tension caused by the hypoosmotic shock or stretching pushed the membrane tension to intermediate or high values in which the coat alone is insufficient to produce closed buds. The observed overabundance of U-shaped, presumably stalled, pits \cite{Boulant2011} is consistent with a situation in which the membrane tension is in the snapthrough regime and the free energy of coat assembly is unable to overcome the energy barrier necessary to deform the membrane into a closed bud shape. Thus, under conditions of hypoosmotic shock it seems that a force exerted by the actin cytoskeleton, as in Figure \ref{fig:pull}, is necessary to overcome the energy barrier required to form a closed bud. 

Avinoam et al. found that the size of the coat does not change substantially during membrane deformation in CME \cite{Avinoam2015}. This is in contrast to the canonical view that the clathrin coat should directly impose its preferred curvature on the underlying membrane \cite{Dannhauser2012}. There are two possible explanations for this observation in the context of our study. One is that the membrane tension is too high for the coat to deform the membrane, so that other mechanisms of curvature generation (e.g. actin polymerization or BAR domain proteins) are necessary to remodel the membrane. The second is that the coat undergoes a ``maturation'' process that progressively increases its capability to bend the membrane, and hence its spontaneous curvature, as in Figure \ref{fig:c0tensionComp}.

In principle, we can distinguish between these two possibilities by measuring the membrane tension in the SK-MEL-2 cell type used in the study. Calculating the vesiculation number could then be used to predict whether the clathrin coat alone is sufficient to produced closed buds or if another membrane remodeling mechanism (e.g. actin assembly) is necessary. Preliminary tether pulling measurements on this cell line yielded tether forces of $38 \pm 12$ pN \textbf{[unpublished data]}, corresponding to a membrane tension of $\lambda_0 = 0.12 \pm 0.08$ pN/nm, assuming a value of $k = 320$ pN$\cdot$nm. This value of membrane tension falls within the high tension or snapthrough regime for a wide range of spontaneous curvatures, indicating that additional facilitators of curvature generation may be necessary to produce closed buds in this cell type. This is consistent with the observation that actin inhibition causes substantial defects in CME in this cell type \cite{Grassart2014}. Thus, it is possible that the findings of Avinoam et al. \cite{Avinoam2015} might be specific to the cell type used and in particular on the typical membrane tension of that cell line.

By reconstituting clathrin coat formation from purified components, Saleem et al. measured the polymerization per unit area of a clathrin coat to be $1.0 \pm 0.5 \times 10^{-4} \,\mathrm{N}\cdot\mathrm{m}^{-1} \approx  2.5 \pm 1.3 \times 10^{-2} \,k_{\mathrm{B}}T \cdot \mathrm{nm}^{-2}$ \cite{Saleem2015}. This energy is sufficient to produce a closed bud at low membrane tensions for the value of the bending modulus, $k = 104 \pm 40$ pN$\cdot$nm, of the synthetic lipid membranes used in this study. This is consistent with the gross membrane deformation and formation of closed buds observed in hypertonic, low membrane tension conditions \cite{Saleem2015}. However, this energy would be insufficient to bridge the energy barrier to coat closing at intermediate membrane tensions. This suggests that this energy barrier might be physiologically relevant. 
Additionally, we can calculate the vesculation number for the membrane tensions ($\approx 0.5 - 3 \,\mathrm{pN}/\mathrm{nm}$) set by micropipette aspiration to be less than 1 over a wide range of spontaneous curvatures, indicating the high membrane tension regime. This is consistent with the shallow buds observed in isotonic conditions.
One result that our model cannot explain is the lack of any clathrin assembly observed under hypotonic conditions \cite{Saleem2015}. It is possible that at extremely high membrane tensions the coat is simply unable to stay bound to the membrane at the extremely flat morphology that would be expected.
Saleem et al. also constructed a theoretical phase diagram comparing the resultant membrane shapes as a function of membrane tension and the polymerization energy of the coat that seems to well explain their observations \cite{Saleem2015}. 
However, their model is limited in that it explicitly assumes that the underlying membrane directly adopts the preferred curvature of the clathrin coat and also that there is a dense packing of buds. Our model does not need to make these assumptions, and we feel that the observations of Saleem et al. can be equally well explained in the context of our modeling framework.

To this point, we have stated that the difference in energy requirement between the open and closed buds in the intermediate tension case is the likely explanation for the appearance of stalled pits in the absence of actin polymerization. An alternative explanation is that clathrin reorganization might be necessary for the clathrin coat to adopt the highly curved, post-snapthrough shape and the coat could be kinetically ``trapped'' in the open morphology.
In order to form closed buds, clathrin triskelia form a meshwork consisting of hexagons and pentagons, similar to a soccer ball \cite{Kirchhausen2000}. A flatter clathrin morphology has relatively more hexagons and so rearrangement of the clathrin lattice would be necessary to facilitate the incorporation of pentagons required for a highly curved morphology \cite{Kirchhausen2014}. This rearrangement would require turnover of the clathrin triskelia as it would be extremely energetically costly to rearrange the lattice while attached to the membrane \cite{Saleem2015}.
Continuous turnover of clathrin at endocytic sites has been observed by flourescence recovery after photobleaching with a half-time of $\approx 2$ seconds \cite{Avinoam2015}. While relatively rapid on the timescale of CME, this might still be too slow to facilitate the snapthrough transition. Each clathrin triskelia transiently lost would decrease the effective size and spontaneous curvature of the coat, and the membrane would then equilibrate to a shape further from the snapthrough transition.
This suggests that a potential role for actin polymerization or BAR-domain proteins could be to maintain membrane curvature during the transition as the clathrin turns over and the lattice reorganizes.

One aspect of CME not explicitly addressed by this study is that the endocytic machinery includes curvature-generating proteins outside of the coat proteins and the actin machinery. In particular, recent modeling studies have demonstrated that cyclindrical curvature induced by BAR-domain proteins can play an important role in reducing the force requirement for productive CME in yeast \cite{Walani2015,Dmitrieff2015}. However, CME is still productive in $50\%$ of events even with complete knockout of the endocytic BAR-domain proteins in this organism \cite{Kishimoto2011}, while actin assembly is absolutely required \cite{Kaksonen2003,Basu2013}. Additionally, in mammalian cells a large percentage of CCPs were found to stall at high membrane tension when actin is inhibited \cite{Boulant2011} despite the fact that the BAR-domain proteins were presumably unaffected. These results suggest that while curvature generated by BAR-domain proteins may help to facilitate productive CME, force from actin assembly seems to be most important in challenging mechanical environments.

Though we have primarily focused on the impact of membrane tension on CME, our findings are general to any budding process. For example, it has been shown that membrane deformation by COPI coats is also inhibited by membrane tension \cite{Manneville2008}. Since the membranes of the endoplasmic reticulum and the Golgi are also under tension \cite{Upadhyaya2004}, one would expect that the shape evolution of buds from these organelles to also be determined by a balance of the coat spontaneous curvature, bending rigidity and membrane tension. Other membrane invaginations are also presumably governed by a similar set of physical parameters. For example, caveolae have been proposed to act as a membrane reservoir that buffers changes in membrane tension by disassembling upon an increase in membrane tension \cite{Sinha2011}. A similar framework to the one used in this study might provide some insight into the morphology and energetics of this process. Moving forward, more detailed measurements of both the membrane tension within cells as well as the spontaneous curvature of various membrane-bending proteins will be essential to verify and extend the results presented here.

%\end{multicols}{2}

%%%%%%%%%%%%%%%%%%%%%%%%%%%%%%%%%%%%%%%%

\section*{Acknowledgements}

The authors would like to thank Matt Akamatsu and Charlotte Kaplan for critical reading of the manuscript. Charlotte Kaplan and Alba Diz-Mu\~{n}oz performed membrane tension measurements referenced in the discussion. This research was supported by a National Defense Science and Engineering Graduate Fellowship to J.E.H, the National Institutes of Health Grant R01GM104979 to G.O. and the UC Berkeley Chancellor's Postdoctoral Fellowship, and the Air Force Office of Scientific Research award number FA9550-15-1-0124 to P.R.

%%%%%%%%%%%%%%%%%%%%%%%%%%%%%%%%%%%%%%%% 

\newpage
\printbibliography

\newpage

\setcounter{equation}{0}
\setcounter{figure}{0}
\setcounter{table}{0}
\setcounter{page}{1}
\renewcommand{\theequation}{S\arabic{equation}}
\renewcommand{\thefigure}{S\arabic{figure}}

\begin{center}
\LARGE Supplemental material for `Membrane tension is a key determinant of bud morphology in clathrin-mediated endocytosis'
\\~\\
\large Julian E. Hassinger$^1$, George Oster$^2$, David G. Drubin$^2$, and $^3$Padmini Rangamani
\\~\\
$^1$Biophysics Graduate Group, University of California, Berkeley

$^2$Department of Molecular and Cell Biology, University of California, Berkeley

$^3$Department of Mechanical and Aerospace Engineering, University of California, San Diego

\end{center}

\tableofcontents

\newpage

%%%%%%%%%%%%%%%%%%%%%%%%%%%%%%%%%%%%%%%%%%%

\section{Model description} \label{sec:extended_model}

\subsection{Assumptions}

\begin{enumerate}

\item The lipid bilayer is modeled as a two-dimensional differentiable manifold endowed with mechanical properties. Helfrich proposed a model that treats the manifold as a thin elastic shell whose bending behavior is captured by an energy density functional that depends only on the manifold's local curvatures \cite{Helfrich1973}. This model and its variants are valid for radii of curvatures much larger than the membrane thickness. We assume that the Helfrich energy is sufficient to describe the membrane during clathrin-mediated endocytosis. 

\item The membrane is assumed to be at mechanical equilibrium at all times. Because clathrin-mediated endocytosis occurs over a timescale of tens of seconds \cite{Taylor2011,Boulant2011,Doyon2011,Aguet2013,Grassart2014}, this assumption is valid \cite{Liu2009, Dmitrieff2015}. We do not include explicit time dependence arising from the viscosity of the bilayer or the surrounding fluid or due to diffusion. This is a focus of future work. 

\item The lipid bilayer is assumed to be incompressible, based on a large stretch modulus \cite{Rawicz2000}. This constraint is introduced using a Lagrange multiplier $\gamma$ (see Table \ref{table:notation1} for notation).

\item Protein adsorption on the membrane (\textit{i.e.} the clathrin coat) is represented using spontaneous curvature ($C$). Since we are modeling the membrane as a single manifold, the notion of  intrinsic curvature due to different lipids or proteins interacting with each leaflet is represented by this term \cite{Helfrich1973,Agrawal2009,Rangamani2014}.

\item For ease of computation, we assume that endocytic pit is rotationally symmetric. This allows us to obtain solutions capturing the whole budding process with a relatively simple parameterization of the surface.

\end{enumerate}

\subsection{Equilibrium equations}

Here we present a concise derivation of the equilibrium equations for biological membranes. Detailed derivations are presented in \cite{Rangamani2013,Steigmann1999,Agrawal2009}.

%\subsubsection{Momentum balance}
%
%The global form of the balance of momentum in the absence of inertia can be written as 
%%
%\begin{equation}\label{eq:globalmomen}
%\frac{d}{dt}\int_{\partial \pi }\fb \, \dm s +  \frac{d}{dt}\int_{\pi }p\mathbf{n} \, \dm a = \zerob,
%\end{equation}
%%
%where $\fb$ is force acting on the boundary of the area $\pi$ and $p$ is the pressure (or body force) distribution on $\pi$. Using general continuumechanical arguments, it can be shown that the force $\fb$ acting on any curved boundary $\partial \pi$ can be written as \cite{Steigmann1999}
%%
%\begin{equation}\label{eq:edge_force_decomp}
%\fb = \sigmab^{\alpha} \nu_{\alpha} 
%\end{equation}
%%
%where $\nu_{\alpha}$ are the components of the normal to the curve $\partial \pi$ in the $\{ \ab^{1},\,\ab^{2}\} $ basis. %
%where $\Tb^{\alpha}$ are components in the tangent plane of the membrane at any point $\rb$ and $S^{\alpha}$ is the normal component. Using \eqref{eq:edge_force_decomp} and the arbitrariness of the surface area $\pi$, 

The local force balance, based on the conservation of linear momentum, and in the absence of inertia is
\begin{equation}\label{eq:mom_bal_local}
\textbf{div} \ \sigmab  +p\nb  =\fb.
\end{equation}
Here, \textbf{div} denotes the surface divergence, $\sigmab$ are the stress vectors, $p$ is the pressure difference across the membrane, $\nb$ is the surface normal, and $\fb$ is the externally applied force. 
The surface stresses can be  expressed as 
\begin{equation}\label{eq:comp_0}
\sigmab^{\alpha} = \Tb^{\alpha} + S^{\alpha}\nb,
\end{equation}
 and the surface divergence is expressed as 

\begin{equation}
 \textbf{div}\ \sigmab = \sigmab^{\alpha}_{;\alpha}=(\sqrt{a})^{-1}(\sqrt{a}\sigmab^{\alpha})_{,\alpha }. 
\end{equation}

\noindent $()_{;\alpha}$ denotes the covariant derivative. 
It should be noted that $\Tb^{\alpha}$ and $S^{\alpha}$ need to be constitutively determined. In this case, if $F(H,K;x^{\alpha})$ is the elastic energy density per unit mass of the surface \cite{Steigmann1999,Rangamani2013}, then $S^{\alpha}$ and the individual components of the $\Tb^{\alpha}$ are given by
\begin{equation}\label{eq:comp_1}
\mathbf{T}^{\alpha }=T^{\beta \alpha }\mathbf{a}_{\beta }\quad \text{%
with\quad }T^{\beta \alpha }=\sigma ^{\beta \alpha }+b_{\mu }^{\beta }M^{\mu
\alpha },\text{\quad and\quad }S^{\alpha }=-M_{;\beta }^{\alpha \beta },
\end{equation}
where
\begin{equation}\label{eq:stress_components}
\sigma ^{\beta \alpha }=\rho (\frac{\partial F}{\partial a_{\alpha \beta }}+%
\frac{\partial F}{\partial a_{\beta \alpha }})\quad \text{and\quad }%
M^{\beta \alpha }=\frac{1}{2}\rho (\frac{\partial F}{\partial b_{\alpha
\beta }}+\frac{\partial F}{\partial b_{\beta \alpha }}) ,
\end{equation}
see \cite{Steigmann1999} for a full derivation. For an elastic membrane that responds to out-of-plane bending and is area incompressible, the general form for the free energy density per unit mass can be rewritten as
\begin{equation}\label{eq:stress_sub_components}
F(\rho ,H,K;x^{\alpha})=\bar{F}(H,K;x^\alpha)-\gamma (x^{\alpha},t)/\rho ,
\end{equation}
where $\gamma(x^{\alpha},t)$ is a Lagrange multiplier required to implement the constraint $\rho(x^\alpha,t)$ is constant or the local area dilation $J=1$, $H$ and $K$ are the mean and Gaussian curvatures respectively. Substituting $W = \rho \bar{F}$ and invoking the definitions of the mean and Gaussian curvatures, $H=\frac{1}{2}a^{\alpha\beta}b_{\alpha\beta}$ and $K=\frac{1}{2}\varepsilon^{\alpha\beta}\varepsilon^{\lambda\mu}b_{\alpha\lambda}b_{\beta}{\mu}$, in terms of the induced metric and curvature tensors $a^{\alpha\beta} = (a_{\alpha\beta})^{-1}$ and $b^{\alpha\beta} = a^{\alpha\lambda}a^{\beta\mu}b_{\lambda\mu}$, respectively, Eq. \eqref{eq:stress_sub_components} can be rewritten as 
\begin{eqnarray}\label{eq:comp_2}
\sigma ^{\alpha \beta } &=&(\lambda +W)a^{\alpha \beta
}-(2HW_{H}+2KW_{K})a^{\alpha \beta }+W_{H}\tilde{b}^{\alpha \beta }, 
\nonumber \\
M^{\alpha \beta } &=&\frac{1}{2}W_{H}a^{\alpha \beta }+W_{K}\tilde{b}%
^{\alpha \beta },  \nonumber \\
\end{eqnarray}
where
\begin{equation}
\lambda =-(\gamma +W).
\end{equation}
Using Eqs. \eqref{eq:comp_2}, \eqref{eq:comp_1} and \eqref{eq:comp_0}, the equations of motion Eq. \eqref{eq:mom_bal_local} can then be reduced to
\begin{equation}\label{eq:elastic_eqn_1}
p + \mathbf{f} \cdot \mathbf{n} = \Delta (\frac{1}{2}W_{H})+(W_{K})_{;\alpha \beta }\tilde{b}^{\alpha \beta
}+W_{H}(2H^{2}-K)+2H(KW_{K}-W)-2\lambda H,
\end{equation}%
and%
\begin{equation}\label{eq:elastic_eqn_2}
N_{;\alpha }^{\beta \alpha }-S^{\alpha }b_{\alpha }^{\beta }=-(\gamma
_{,\alpha }+W_{K}K_{,\alpha }+W_{H}H_{,\alpha })a^{\beta \alpha }=(\partial
W/\partial x _{\mid \exp }^{\alpha }+\lambda _{,\alpha })a^{\beta
\alpha} = 0,
\end{equation}%

Here $\Delta(\cdot)=(\cdot)_{;\alpha\beta}a^{\alpha\beta}$ is the surface Laplacian and $()_{\mid \exp}$ represents the explicit derivative with respect to $\theta^\alpha$. 

\subsubsection{Helfrich energy elastic model}

For a lipid-bilayer membrane, the local elastic energy takes the Helfrich energy form 

\begin{align} \label{eq:Helfrich}
W = k \left( H - C(\theta^\alpha) \right)^2 + \bar{k} K
\end{align}

\noindent where $C \left(\theta^{\alpha} \right)$ is the spontaneous curvature that can depend on the coordinates. This energy function differs from the standard Helfrich energy by a factor of $2$, with the net effect being that our value for the bending modulus, $k$, is twice that of the standard bending modulus typically encountered in the literature \cite{Helfrich1973}. The Gaussian modulus is assumed to be uniform, and the membrane is planar at the boundary of the simulated domain. 

The equations of motion \eqref{eq:elastic_eqn_1} and \eqref{eq:elastic_eqn_2} for an elastic membrane reduce to 

\begin{align} \label{eq:shape}
k \Delta \left(H - C \right) + 2 k \left( H - C \right) \left(2 H^2 - K \right) - 2 k H \left( H - C \right)^2 = p + 2 \lambda H + \mathbf{f} \cdot \mathbf{n},
\end{align}

\begin{align} \label{eq:lambda}
\lambda_{, \gamma} = -\frac{\partial{W}}{\partial{x^{\gamma}}} = 2 k \left( H - C \right) \frac{\partial{C}}{\partial{x^{\gamma}}} - \mathbf{f} \cdot \mathbf{a}_{\gamma}.
\end{align}

In the absence of externally applied force $\fb$, we recover the equations of motion for a heterogeneous membrane \cite{Agrawal2009,Rangamani2014}.

\begin{equation}
\label{eq:elastic_flow_app}
\lambda _{,\gamma }= -\frac{\partial W}{\partial x^{\gamma}}{\mid_{\exp}} =2k(H-C)\frac{\partial C}{\partial x^{\gamma}}.
\end{equation}
\begin{eqnarray}
\label{eq:elastic_shape_app}
k[\Delta (H-C)+2(H-C)(H^2+HC-K)]-2\lambda H=p.
\end{eqnarray}

 $\lambda$ can be interpreted as the tension in a flat membrane \cite{Rangamani2013,Steigmann1999}. Furthermore, in the special case of zero spontaneous curvature and non-zero mean curvature, $\lambda=constant$ everywhere, see Eq. \eqref{eq:lambda}. This constant value of $\lambda$ must be provided as an input parameter to solve the system of equations \cite{Steigmann1999}, and is widely interpreted to be surface tension in the literature \cite{Derenyi2002}. A detailed interpretation of $\lambda$ is given in \cite{Rangamani2014}.

\subsection{Equations of motion in axisymmetric coordinates} \label{ss:axi}

\subsubsection{Definitions}

We define a surface of revolution (see Figure \ref{fig:geometry})

\begin{align} \label{eq:vecr}
\mathbf{r}\left( s, \theta \right) = r(s) \mathbf{e}_r \left( \theta \right) + z(s) \mathbf{k}
\end{align}

\noindent where $s$ is the arc-length along the curve, $r(s)$ is the radius from axis of revolution, $z(s)$ is the elevation from a base plane and $(\mathbf{e}_r, \mathbf{e}_{\theta}, \mathbf{k})$ form the coordinate basis. Since $(r')^2 + (z')^2 = 1$, we can define an angle $\psi$ such that

\begin{align} \label{eq:bases}
\mathbf{a}_s = \cos{\psi} \mathbf{e}_r + \sin{\psi} \mathbf{k}, \quad \mathbf{n} = - \sin{\psi} \mathbf{e}_r + \cos{\psi} \mathbf{k}
\end{align}

\noindent are the unit tangent and normal vectors, respectively, and

\begin{align} \label{eq:rzs}
r'(s) = \cos{\psi}, \quad  z'(s) = \sin{\psi}
\end{align}

\noindent parametrize the surface. The tangential ($\kappa_{\nu}$) and transverse ($\kappa_{t}$) curvatures are given by

\begin{align} \label{eq:axicurv}
\kappa_{\nu} = \psi, \quad \kappa_{t} = r^{-1} \sin{\psi}
\end{align}

\noindent and the mean ($H$) and Gaussian ($K$) curvatures 

\begin{align} 
H &= \frac{1}{2} \left( \kappa_{\nu} + \kappa_{t} \right) \label{eq:mean} \\
K &= \kappa_{\nu} \kappa_{t} = H^2 - \left(H - r^{-1} \sin{\psi} \right)^2. \label{eq:gauss}
\end{align}

Rearranging Eq. \eqref{eq:mean} with Eq. \eqref{eq:axicurv} yields the differential equation for $\psi$,

\begin{align} \label{eq:psis}
r \psi'(s) = 2 r H - \sin{\psi} .
\end{align}

\subsubsection{Equilibrium equations}

Let $L$ be given by

\begin{align} \label{eq:defL}
L = \frac{1}{2 k} r \left( W_H \right)' = r \left( H - C \right)',
\end{align}

\noindent allowing us to obtain a first-order differential equation for the mean curvature,

\begin{align} \label{eq:Hs}
H'(s) = r^{-1} L + C'(s).
\end{align}

Using Eq. \eqref{eq:defL} in  Eq. \eqref{eq:shape} we obtain

\begin{align} \label{eq:delH}
k \Delta \left(H -C \right) = k \left[ \left( H - C \right)' \right]' = k r^{-1} \left[ L \right]' = k r^{-1} L'.
\end{align}

Inserting Eq.  \eqref{eq:delH} into Eq. \eqref{eq:shape} and rearranging we obtain a first-order differential equation for L,

\begin{align} \label{eq:Ls}
k r^{-1} L' &+ 2 k \left(H - C \right) \left(2 H^2 - K \right) - 2 k H \left(H - C \right)^2 = p + 2 \lambda H + \mathbf{f} \cdot \mathbf{n} \notag \\
k r^{-1} L' &= p + \mathbf{f} \cdot \mathbf{n} + 2 H \left[ k \left(H - C \right)^2 + \lambda \right] - 2 k \left(H - C \right) \left[ H^2 + \left( H - r^{-1} \sin{\psi} \right)^2 \right] \notag \\
r^{-1} L'(s) &= \frac{p}{k} + \frac{\mathbf{f} \cdot \mathbf{n}}{k} + 2 H \left[ \left(H - C \right)^2 + \frac{\lambda}{k} \right] - 2 \left(H - C \right) \left[ H^2 + \left( H - r^{-1} \sin{\psi} \right)^2 \right].
\end{align}

Finally, Eq. \eqref{eq:lambda} becomes

\begin{align} \label{eq:lams}
\lambda'(s) = 2 k \left( H - C \right) C'(s) - \mathbf{f} \cdot \mathbf{a}_{s}.
\end{align}

The system of equations to be solved to obtain the shapes of the membrane are Eqs. \eqref{eq:rzs}, \eqref{eq:psis}, \eqref{eq:Hs}, \eqref{eq:Ls}, and \eqref{eq:lams}.

\subsubsection{Boundary conditions} \label{sss:bc}

\begin{figure}[h]
\centering
\includegraphics[width=0.95\linewidth, height = 0.7\textheight, keepaspectratio]{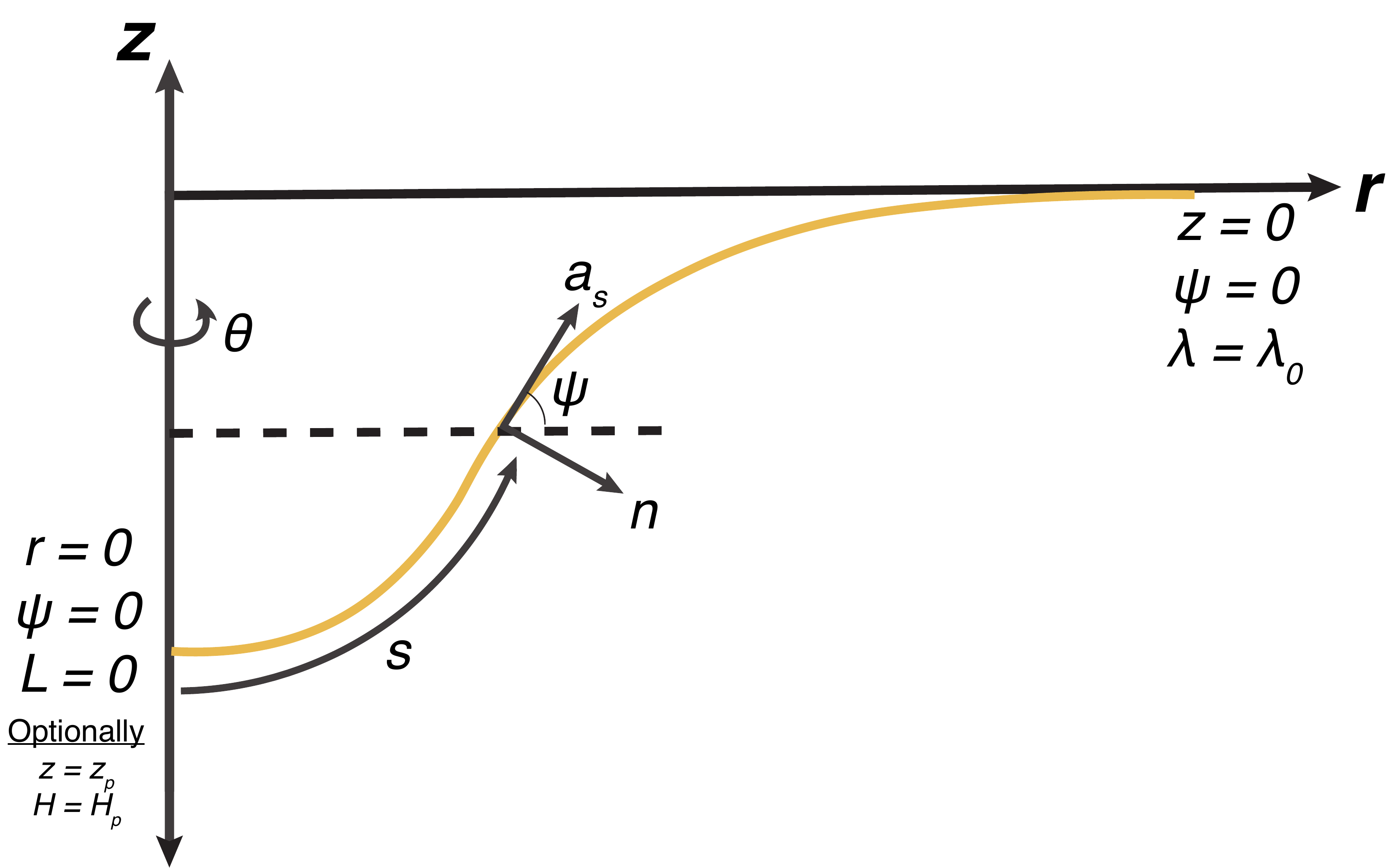}

\caption{Schematic of the axisymmetric geometry adopted for the simulations as described in Section \ref{ss:axi}. The boundary conditions at the tip of the bud and the boundary of the patch were implemented as indicated. The optional boundary conditions \eqref{eq:Zp} and \eqref{eq:Hp} were used to obtain the value of applied force in actin-mediated pulling (Figure \ref{fig:pull}) and pinching (Figure \ref{fig:squeeze}) simulations, respectively (see Section \ref{sss:bc}).
}

\label{fig:geometry}

\end{figure}

In order to solve this system of equations, we need to provide six boundary conditions. We consider an axisymmetric circular patch of membrane (see Figure \ref{fig:geometry}). At the center of the patch, $s = 0^+$, we require: 1) the distance from the axis of symmetry be 0, 2) $\psi = 0$ to ensure continuous differentiability of the surface, and 3) $L = 0$ due to reflection symmetry. At the boundary of the patch, $s = S$, we require 1) that the membrane not lift off and therefore $Z = 0$, 2) $\psi = 0$ to ensure continuous differentiability with the flat surrounding membrane, and 3) $\lambda$ is prescribed. These conditions can be summarized as 

\begin{subequations} \label{eq:bc}
\begin{align}
R\left(0^{+}\right) = 0,\quad  L \left( 0^{+} \right) = 0, \quad  \psi\left( 0^{+} \right) = 0, \\
\quad Z\left(S\right) = 0, \quad \psi\left( S \right) = 0, \quad \lambda\left( S \right) = \lambda_0.
\end{align}
\end{subequations}

In cases relating to the actin-mediated pulling (Figure \ref{fig:pull}), we prescribe the displacement of the of the tip, and calculate the force needed to maintain the prescribed displacement. We implement this additional boundary condition as 

\begin{equation} \label{eq:Zp}
Z \left( 0^{+} \right) = Z_p.
\end{equation}

Similarly, for actin-mediated pinching (Figure \ref{fig:squeeze}), we prescribe the mean curvature at the bud tip and calculate the force required to maintain this curvature. This boundary condition is implemented as

\begin{equation} \label{eq:Hp}
H \left( 0^{+} \right) = H_p.
\end{equation}

\subsubsection{Dimensionless variables}

In order  to perform the numerical computations, we non-dimensionalized the system by introducing two positive constants, $R_0$ and $k_0$, and defining the following dimensionless variables.

\begin{align} \label{eq:arcdldef}
\begin{split}
t \equiv s/R_0, \quad x \equiv r/R_0, \quad y \equiv z/R_0, \quad h \equiv H R_0, \quad c \equiv C R_0, \\
l \equiv L R_0, \quad \widetilde{\lambda} \equiv \lambda {R_0}^2/k_0, \quad \widetilde{p} \equiv p {R_0}^3/k_0, \quad \widetilde{\mathbf{f}} \equiv \mathbf{f} {R_0}^3/k_0, \quad \widetilde{k} \equiv k/k_0.
\end{split}
\end{align}

In terms of Eq. \eqref{eq:arcdldef}, the system of equations, Eqs. \eqref{eq:rzs}, \eqref{eq:psis}, \eqref{eq:Hs}, \eqref{eq:Ls}, and \eqref{eq:lams}, become

\begin{subequations} \label{eq:arcdleqs}
\begin{align} 
\dot{x} = \cos{\psi}, \quad \dot{y} = \sin{\psi}, \quad x \dot{\psi} = 2 x h - \sin{\psi}, \quad \dot{h} = x^{-1} l + \dot{c} \\
x^{-1} \dot{l} = \frac{\widetilde{p}}{\widetilde{k}} + \frac{\widetilde{\mathbf{f}} \cdot \mathbf{n}}{\widetilde{k}} + 2 h \left[ \left(h - c \right)^2 + \frac{\widetilde{\lambda}}{\widetilde{k}} \right] - 2 \left( h - c \right) \left[ h^2 + \left( h - x^{-1} \sin{\psi} \right)^2 \right] \\
\dot{\widetilde{\lambda}} = 2 \widetilde{k} \left( h - c \right) \dot{c} - \widetilde{\mathbf{f}} \cdot \mathbf{a_s}
\end{align}
\end{subequations}

the boundary conditions, Eq. \eqref{eq:bc} become

\begin{subequations} \label{eq:arcdlbc}
\begin{align}
x\left(0^{+}\right) = 0, \quad l \left( 0^{+} \right) = 0, \quad \psi\left( 0^{+} \right) = 0 \\
\psi\left( T \right) = 0, \quad y\left(T\right) = 0, \quad \widetilde{\lambda}\left( T \right) = \widetilde{\lambda}_0,
\end{align}
\end{subequations}

where $T = S/R_0$ is the total dimensionless arc-length. The boundary conditions imposed to solve for unknown applied force, Eqs. \eqref{eq:Zp} and \eqref{eq:Hp}, become

\begin{equation} \label{eq:arcyphp}
y\left( 0^{+} \right) = y_p, \quad h\left( 0^{+} \right) = h_p.
\end{equation}

\subsection{Area dependence}

\subsubsection{Arc-length to area dependence}
In axisymmetry there is a one-to-one correspondence between arc-length and area, which allows us to express the system of equations as a function of area instead of arc-length. This is because 

\begin{align} \label{eq:arc2area}
a(s) = 2 \pi \int_0^s r(t) \D{t}  \quad \implies \quad \frac{\D{a}}{\D{s}} =  2 \pi r.
\end{align}
This method has the advantage of prescribing the total membrane area rather than the arc length as the simulation domain. 
Using Eq. \eqref{eq:arc2area}, we convert Eqs. \eqref{eq:rzs}, \eqref{eq:psis}, \eqref{eq:Hs}, \eqref{eq:Ls}, and \eqref{eq:lams} into

\begin{subequations} \label{eq:areaeqs}
\begin{align} 
2 \pi r \,r'(a) = \cos{\psi}, \quad 2 \pi r\, z'(a) = \sin{\psi}, \quad 2 \pi r^2 \psi'(a) = 2 r H - \sin{\psi}, \\
2 \pi r^2 H'(a) = L + 2 \pi r^2 C'(a) \\
2 \pi L'(a) = \frac{p}{k} + \frac{\mathbf{f}\cdot \mathbf{n}}{k} + 2 H \left[ \left( H - C \right)^2 + \frac{\lambda}{k} \right] - 2 \left( H - C \right)\left[ H^2 + \left( H - r^{-1} \sin{\psi} \right)^2 \right] \\
2 \pi r \lambda'(a) = 4 \pi r k \left( H - C \right) C'(a) - \mathbf{f}\cdot\mathbf{a_s},
\end{align}
\end{subequations}

The boundary conditions, Eqs. \eqref{eq:bc}  and \eqref{eq:Zp} are remain unchanged except for the limits at which they are applied. 

\begin{subequations} \label{eq:areabc}
\begin{align}
R\left(0^{+}\right)= 0, \quad L \left( 0^{+} \right) = 0, \quad \psi\left( 0^{+} \right) = 0, \\
\psi\left( A \right) = 0, \quad Z\left(A \right) = 0, \quad \lambda\left( A \right) = \lambda_0,
\end{align}
\end{subequations}
where $A = 2 \pi \int_0^S r(t) \D{t}$ is the total area of the membrane patch. The displacement and mean curvature conditions are
\begin{equation} \label{eq:areaZpHp}
Z \left( 0^{+} \right) = Z_p, \quad H \left( 0^{+} \right) = H_p.
\end{equation}

\subsubsection{Dimensionless variables}
For Figures \ref{fig:tensionComp}, \ref{fig:c0tensionComp}, \ref{fig:endoCurv} and \ref{fig:pull}, we used non-dimensional area dependent equations. Using the two positive constants, $R_0$ and $k_0$, we can define

\begin{align} \label{eq:areadldef}
\begin{split}
\alpha \equiv \frac{a}{2 \pi {R_0}^2}, \quad x \equiv r/R_0, \quad y \equiv z/R_0, \quad h \equiv H R_0, \quad c \equiv C R_0, \\
l \equiv L R_0, \quad \widetilde{\lambda} \equiv \lambda {R_0}^2/k_0, \quad \widetilde{p} \equiv p {R_0}^3/k_0, \quad \widetilde{\mathbf{f}} \equiv \mathbf{f} {R_0}^3/k_0, \quad \widetilde{k} \equiv k/k_0.
\end{split}
\end{align}

In terms of Eq. \eqref{eq:areadldef}, the system of equations, Eq. \eqref{eq:areaeqs}, become

\begin{subequations} \label{eq:areadleqs}
\begin{align} 
x \dot{x} = \cos{\psi}, \quad x \dot{y} = \sin{\psi} \quad x^2 \dot{\psi} = 2 x h - \sin{\psi}, \quad x^2 \dot{h} &= l + x^2 \dot{c} \\
\dot{l} = \frac{\widetilde{p}}{\widetilde{k}} + \frac{\widetilde{\mathbf{f}} \cdot \mathbf{n}}{\widetilde{k}} + 2 h \left[ \left(h - c \right)^2 + \frac{\widetilde{\lambda}}{\widetilde{k}} \right] - 2 \left( h - c \right) \left[ h^2 + \left( h - x^{-1} \sin{\psi} \right)^2 \right] \\
\dot{\widetilde{\lambda}} = 2 \widetilde{k} \left( h - c \right) \dot{c} - x^{-1} \widetilde{\mathbf{f}} \cdot \mathbf{a_s},
\end{align}
\end{subequations}

and the boundary conditions, Eq. \eqref{eq:areabc} and Eq, \eqref{eq:areaZpHp}, become

\begin{subequations} \label{eq:areadlbc}
\begin{align}
x\left(0^{+}\right) = 0, \quad l \left( 0^{+} \right) = 0, \quad \psi\left( 0^{+} \right) = 0 \\
y\left(\alpha_{max} \right) = 0 ,\quad \psi\left( \alpha_{max} \right) = 0, \ \widetilde{\lambda}\left(\alpha_{max} \right) = \widetilde{\lambda}_0,\\
y\left( 0^{+} \right) = y_p \quad h\left( 0^{+} \right) = h_p.
\end{align}
\end{subequations}

where $\alpha_{max} = \frac{A}{2 \pi {R_0}^2}$ is the total dimensionless membrane area.

%%%%%%%%%%%%%%%%%%%%%%%%%%%%%%%%%%%%%%%%%%%
\section{Simulation Methods} \label{sec:sim}

Computations were performed using MATLAB\textsuperscript{\textregistered} (Mathworks, Natick, MA)  using the routine `bvp4c', a boundary value problem solver. All MATLAB\textsuperscript{\textregistered} code used to perform the simulations can be found online as Supplemental files.

\begin{figure}[h]

\includegraphics[width=0.95\linewidth, height = 0.7\textheight, keepaspectratio]{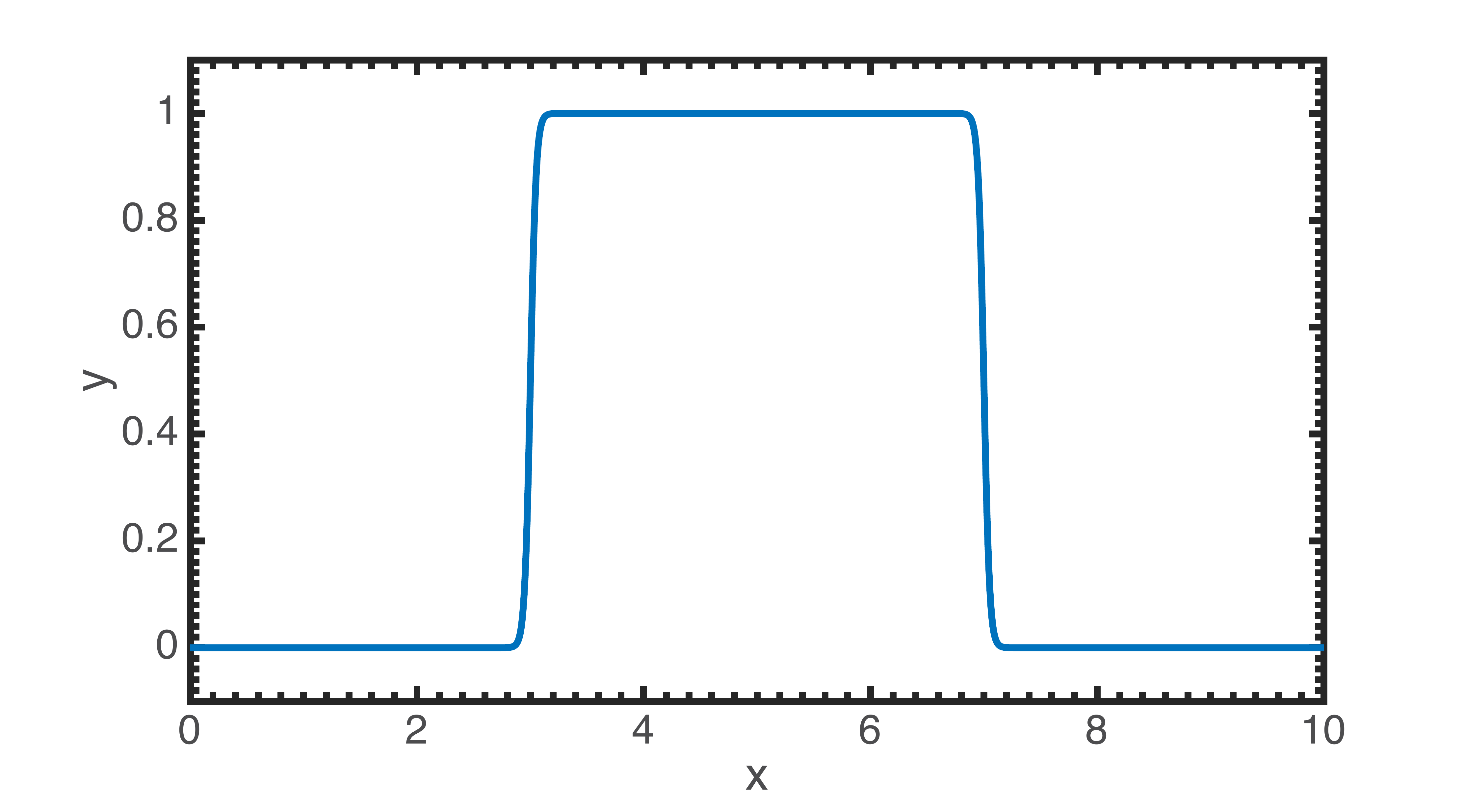}

\caption{A hyperbolic tangent functional form was used to implement heterogeneous membrane properties. As an example, $y = \frac{1}{2}\left[\tanh(\gamma(x-3)) - \tanh(\gamma(x-7))\right]$ is plotted with $\gamma = 20$. The sharp transitions were ideal for specifying the boundaries of the coated region or regions of applied force, and the smoothness of the $\tanh$ function allowed for straightforward implementation into the numerical scheme.
}

\label{fig:tanh}

\end{figure}

\begin{figure}[h]

\includegraphics[width=0.95\linewidth, height = 0.7\textheight, keepaspectratio]{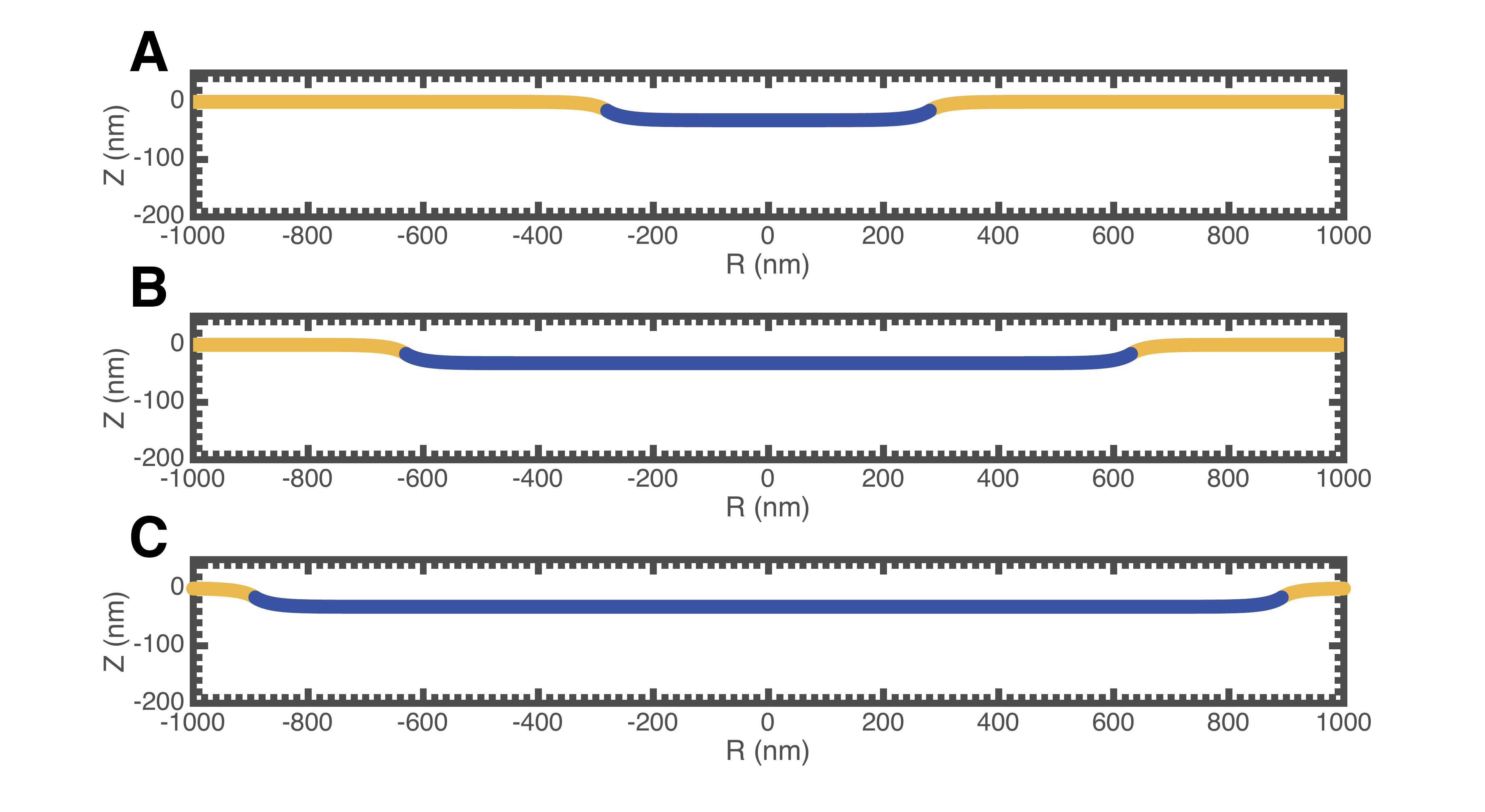}

\caption{High membrane tension, $\lambda_0 = 0.2 \, \textrm{pN}/\textrm{nm}$, $C_0 = 0.02 \,\mathrm{nm}^{-1}$. At high membrane tension, the coat can grow arbitrarily large without causing a substantial deformation of the membrane. \textbf{(A)} $A_{\mathrm{coat}} = 251{,}327 \, \mathrm{nm}^2$, \textbf{(B)} $A_{\mathrm{coat}} = 1{,}256{,}637 \, \mathrm{nm}^2$, \textbf{(C)} $A_{\mathrm{coat}} = 2{,}513{,}274 \, \mathrm{nm}^2$}

\label{fig:highTensionBigCoat}

\end{figure}

\begin{figure}[h]
\centering
\includegraphics[width=0.95\linewidth, height = 0.7\textheight, keepaspectratio]{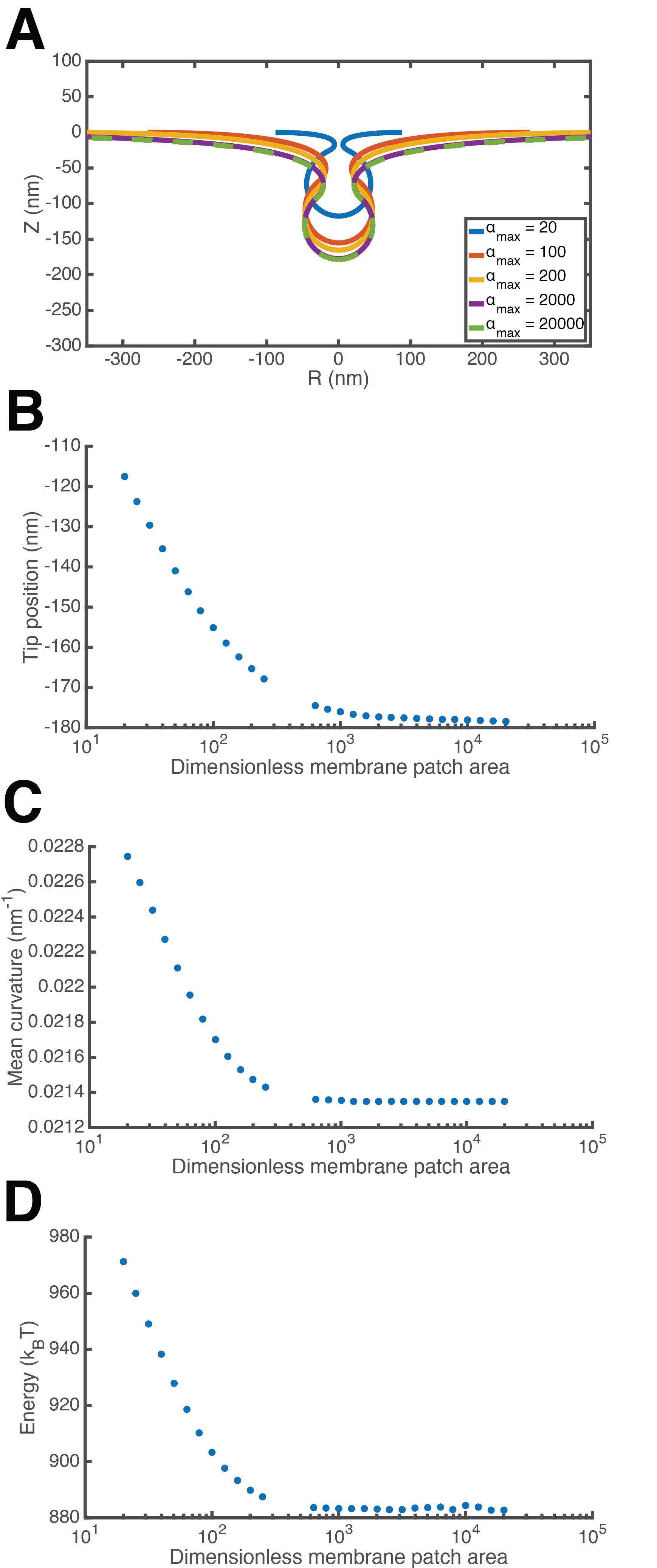}

\caption{The size of the membrane patch has essentially no effect on the observed deformations of the membrane as long as it is sufficiently large. \textbf{(A)} Membrane profiles for identical coat areas and differing total patch areas. $\lambda_0 = 0.002 \,\mathrm{pN}/\mathrm{nm}$, $A_{\mathrm{coat}} = 25,133 \,\mathrm{nm}^2$, $C_0 = 0.02 \,\mathrm{nm}^{-1}$. The deformations are identical for very large membrane patches. \textbf{(B-D)} Z-position of the bud tip, mean curvature of the bud tip, and energy to deform the membrane, respectively, as a function of the dimensionless area of the membrane patch. The deformation of the membrane is sensitive to small membrane patches, but is essentially identical beyond $\alpha_{\mathrm{max}} \approx 200$, particularly in terms of the tip mean curvature and the deformation energy.}

\label{fig:alphaComp}

\end{figure}

\begin{figure}
\centering
\includegraphics[width=0.95\linewidth, height = 0.7\textheight, keepaspectratio]{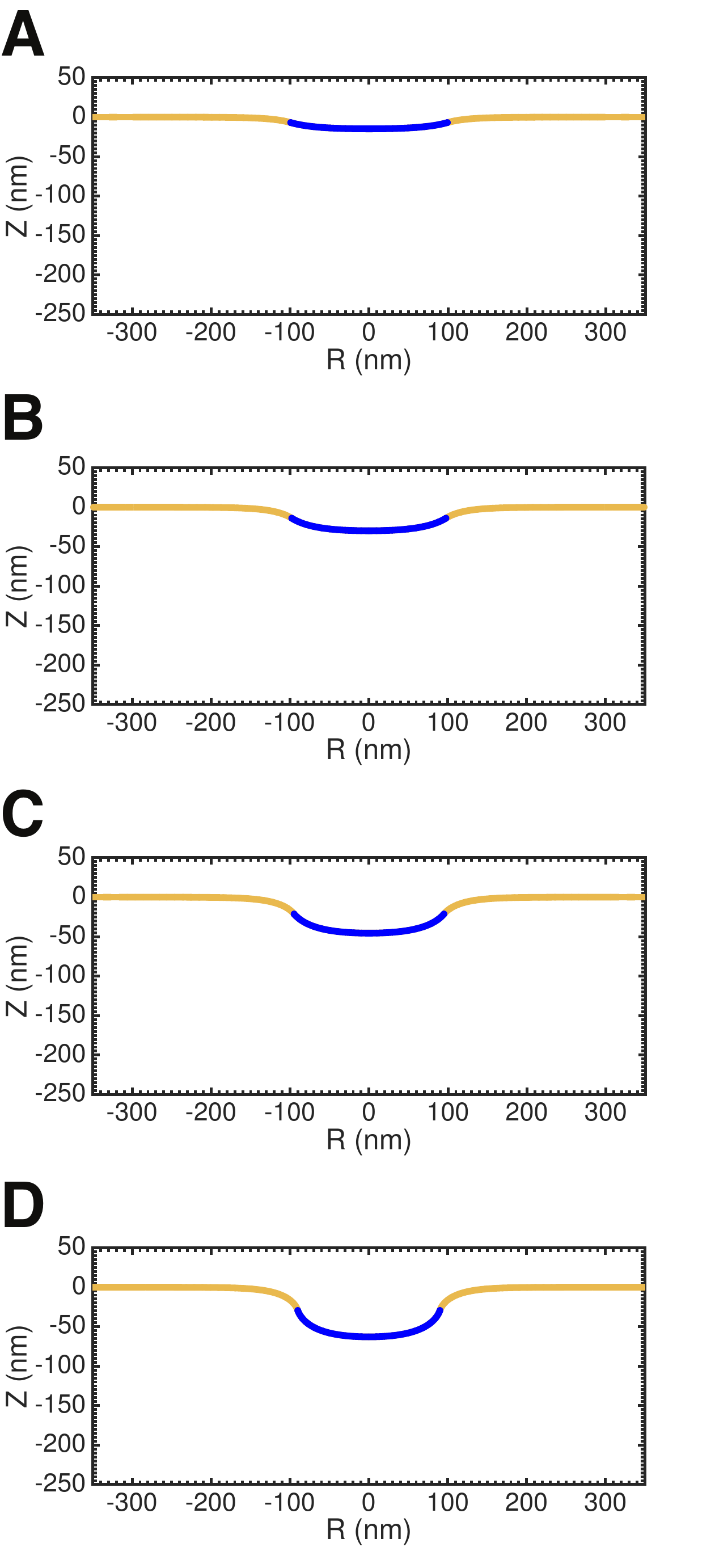}

\caption{Increasing the spontaneous curvature of the coat at high membrane tensions does not produce closed buds. $A_{\mathrm{coat}} = 31,416 \,\textrm{nm}^2$, $\lambda_0 = 0.2 \, \textrm{pN}/\textrm{nm}.$ \textbf{(A)} $C_0 = 0.01 \, \mathrm{nm}^{-1}$ \textbf{(B)} $C_0 = 0.02 \, \mathrm{nm}^{-1}$ \textbf{(C)} $C_0 = 0.03 \, \mathrm{nm}^{-1}$ \textbf{(D)} $C_0 = 0.04 \, \mathrm{nm}^{-1}$}

\label{fig:endoCurvHighTension}

\end{figure}

\begin{figure}
\centering
\includegraphics[width=0.95\linewidth, height = 0.7\textheight, keepaspectratio]{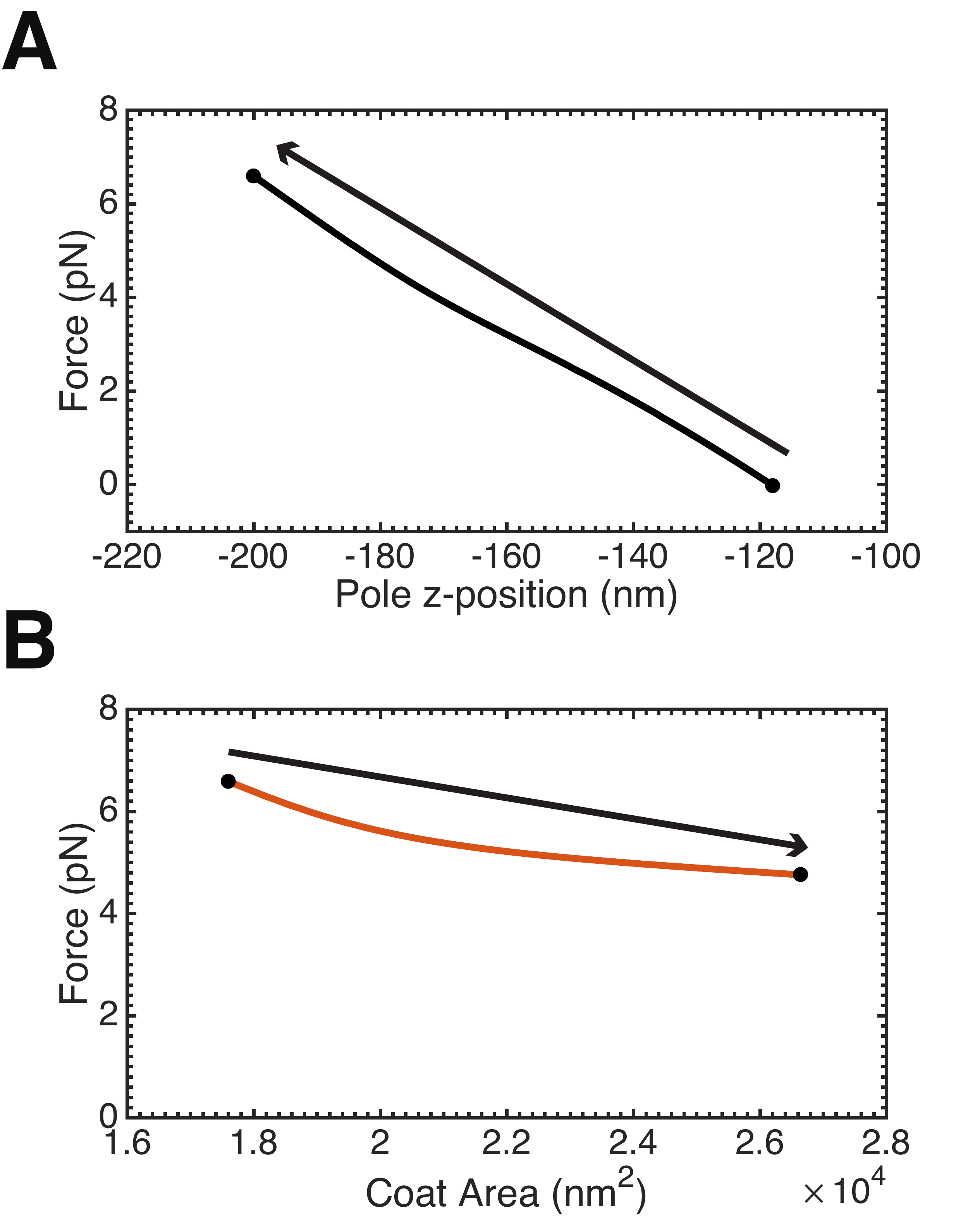}

\caption{Pulling force simulations. \textbf{(A)} Applied force as a function of the z-position of the bud tip. The arrow denotes the direction of the simulation with the marked points denoting the solutions depicted in Figure \ref{fig:pull}A. The applied force necessary to hold the bud at the prescribed depth increases approximately linearly with increasing depth. \textbf{(B)} Applied force as a function of the area of the coat. The arrow denotes the direction of the simulation with the marked points denoting the solutions depicted in Figure \ref{fig:pull}B. The force necessary to hold the bud tip at $Z = -200 \,\mathrm{nm}$ decreases slightly with increasing coat area.}

\label{fig:FvsZpA0}

\end{figure}

\begin{figure}[h]
\centering
\includegraphics[width=0.95\linewidth, height = 0.7\textheight, keepaspectratio]{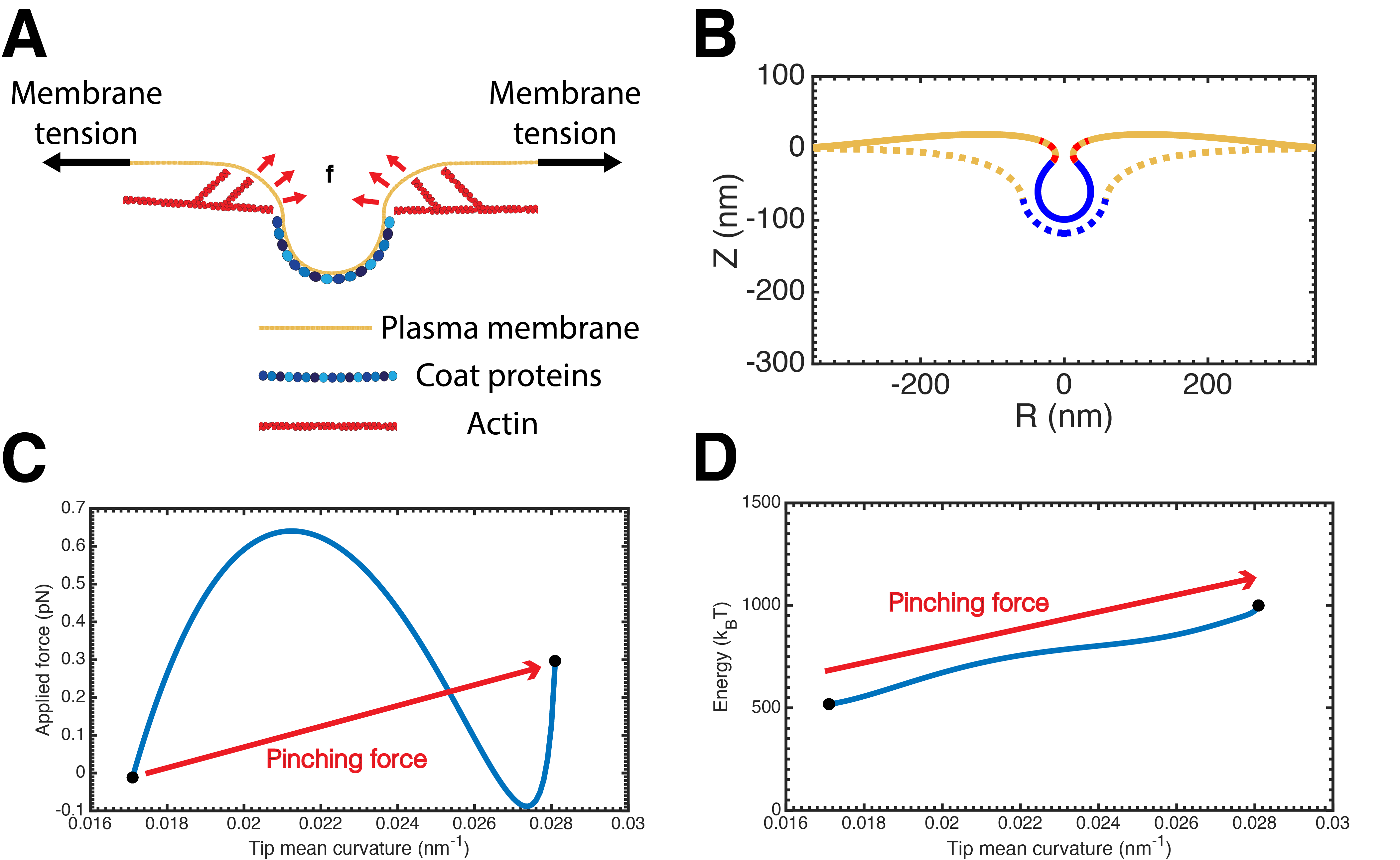}

\caption{
A pinching force can mediate the transition from a U- to $\Omega$-shaped bud. The actin force is oriented normal to the membrane in a ``collar'' situated immediately next to the coated region. \textbf{(A)} Schematic showing the orientation and location of the pinching force. \textbf{(B)} Profile view showing the shape of the membrane before (dashed line) and after (solid line) application of a pinching force at constant coat area, $A_{\mathrm{coat}}= 17,593 \,\mathrm{nm}^2$. \textbf{(C)} Total applied force as a function of the mean curvature at the tip of the bud. The force remains below $1 \,\mathrm{pN}$, well within the capability of a few actin filaments \cite{Bieling2016}. \textbf{(D)} Energy necessary to deform the membrane as a function of the mean curvature at the tip of the bud. Approximately $500 \,k_{\mathrm{B}}T$ is necessary to deform the membrane from the open U-shape to the closed $\Omega$-shape as depicted in (B), well within the capability of a few hundred actin monomers assuming a $\approx 5\%$ energy efficiency \cite{Bieling2016}. 
}

\label{fig:squeeze}

\end{figure}

\begin{itemize}
\item The membrane patch is initialized to be a flat disk with a radius of $R = 400 \,\mathrm{nm}$.

\begin{itemize}
\item The exception to this was for the simulations involving the creation of the phase diagram, Figure \ref{fig:phase}. In this case, the initial radius of the disk was set such that the arc-length of the domain was twice that needed for a bud of radius $R = 1/C_0$, where $C_0$ is the spontaneous curvature of the coat.
\end{itemize}

\item The mesh points on the domain were chosen such that they were (initially) equally spaced along the arc-length with a spacing of $0.5 \,\mathrm{nm}$. To obtain convergence, the solver was allowed to increase the number of mesh points by up to a factor of 100, with the final solution evaluated on the original mesh.

\item Subsequently, the area (or arc-length) of the coat (with fixed spontaneous curvature) or the spontaneous curvature of the coat (with fixed coat area) were progressively increased, with each solution in this sequence used in the solver as the initial guess for the subsequent computation.

\item To ensure sharp but smooth transitions at the boundaries of the coat and regions of applied force, these regions were specified using a hyperbolic tangent function (Figure \ref{fig:tanh}).

\item Simulations using the area-dependent equations \eqref{eq:areadleqs}: Figures \ref{fig:tensionComp}, \ref{fig:c0tensionComp}, \ref{fig:endoCurv}, \ref{fig:pull}, \ref{fig:highTensionBigCoat}, and \ref{fig:endoCurvHighTension}

\begin{itemize}
\item This was the preferred method for solving the ODEs as the area of the membrane domain is kept constant throughout the entire simulation, meaning that the boundaries of the domain stayed relatively more constant than in the arc-length parametrization. 
\item Despite this preference, we show in Figure \ref{fig:alphaComp} that the area of the membrane patch makes essentially no difference on the observed shapes of the membrane as long as the domain is sufficiently large, ensuring that either surface parametrization is valid and the solutions directly comparable.
\item Additionally, this parametrization is much more convenient in terms of directly specifying the coat area as well as the area of applied force. The applied force case is especially important as the applied force is really a force per unit area, or effective pressure (see Eq. \eqref{eq:shape}), and the total magnitude of the force is obtained by integrating the force per unit area over the applied area. 
\end{itemize}

\item Simulations using the arc-length-dependent equations \eqref{eq:arcdleqs}: Figures \ref{fig:snapthrough} and \ref{fig:phase}

\begin{itemize}
\item Arc-length dependence was convenient for ``coat-growing'' simulations in the case of intermediate membrane tension as this formulation allowed for the instability to be smoothly traversed by simply increasing the arc-length coverage by the coat.
\item This is because it is possible for the arc-length covered by the coat to increase while the area coverage of the coat simultaneously decreases if the membrane becomes much more highly curved in the coated region.
\item This is exactly what happens for the so-called \emph{unstable solutions} in Figure \ref{fig:snapthrough}. These solutions cannot be accessed by the solver simply by adjusting the area of the coat; the next solution will always fall on one of the two \emph{stable solution} branches. However, these unstable solutions are readily accessible simply by increasing the arc-length coverage of the coat, conveniently tracing out the entire solution space.
\end{itemize}

\item Simulations with an applied force: Figures \ref{fig:pull}, \ref{fig:FvsZpA0}, and \ref{fig:squeeze}
\begin{itemize}
\item Rather than prescribing the force to obtain a displacement, we took advantage of the ability of the `bvp4c' solver to calculate values of unknown parameters given an initial guess and an additional boundary condition.
\item For pulling forces (Figures \ref{fig:pull} and \ref{fig:FvsZpA0}), this was achieved by specifying the z-position of the tip of the bud Eq. \eqref{eq:areaZpHp}. For simplicity, the force was applied uniformly across the coat.
\item For squeezing forces (Figure \ref{fig:squeeze}), this was achieved by specifying the mean curvature at the tip of the bud Eq. \eqref{eq:areaZpHp}. For simplicity, the applied force was applied on a band of the membrane immediately bordering the coat.
\end{itemize}

\end{itemize}

%%%%%%%%%%%%%%%%%%%%%%%%%%%%%%%%%%%%%%%%%%%

\section{Computation of the energy to deform the membrane} \label{s:energy}

\begin{itemize}
\item The energy necessary to deform the membrane can be expressed as the sum of the work done against the bending rigidity, membrane tension, and pressure: 
\begin{align} \label{eq:Etot}
E_{\mathrm{tot}} = E_{\mathrm{bending}} + E_{\mathrm{tension}} + E_{\mathrm{pressure}}
\end{align}

\item The bending energy is calculated by integrating the bending rigidity, $k$, times the square of the difference between the local mean curvature, $H$, and the local spontaneous curvature, $C$, over the area of the domain:
\begin{align} \label{eq:Ebend}
E_{\mathrm{bending}} = \int_{0}^{A} k \left(H - C \right)^2 \D a
\end{align}

\begin{itemize}

\item For comparison, this is equivalent to integrating Eq. \eqref{eq:Helfrich} over the area of the domain.

\end{itemize}

\item The work done against membrane tension is calculated by multiplying the edge membrane tension, $\lambda_0$, by the difference between the total and projected area of the domain:
\begin{align} \label{eq:Etension}
E_{\mathrm{tension}} = \lambda_0 \left(A_{\mathrm{tot}} - A_{\mathrm{proj}} \right)
\end{align}

\begin{itemize}

\item $\left( A_{\mathrm{tot}} - A_{\mathrm{proj}} \right)$ represents the amount of area that must be pulled in from the surrounding membrane to accommodate the deformation of the membrane patch. Multiplication by the edge membrane tension (the value of the membrane tension in the surrounding membrane) gives the energy of the deformation.

\item The projected area is simply the area of the disk obtained when the membrane deformation is projected onto the plane $Z = 0 \,\mathrm{nm}$. This quantity is easily calculated as $A_{\mathrm{proj}} = \pi {R_{\mathrm{bound}}}^2$, where $R_{\mathrm{bound}}$ is the radial distance to the boundary.

\item In the case of arc-length dependence, $A_{\mathrm{tot}} = 2 \pi \int_0^S r(t) \D{t}$, where $S$ is the total arc-length. In the case of area dependence $A_{\mathrm{tot}}$ is simply the set value of the area of the domain.

\end{itemize}

\item The work done against pressure is calculated by multiplying the transmembrane pressure, $p$, by the volume enclosed by the deformed membrane and the plane $Z = 0 \,\mathrm{nm}$, $V_{\mathrm{encl}}$:
\begin{align} \label{eq:Epressure}
E_{\mathrm{pressure}} = p V_{\mathrm{encl}}
\end{align}

\begin{itemize}

\item The enclosed volume is calculated as $V_{\mathrm{encl}} = \pi \int_{Z(a=0)}^{Z(a=A)} R(a)^2 \D{Z(a)}$.

\end{itemize}

\end{itemize}

%%%%%%%%%%%%%%%%%%%%%%%%%%%%%%%%%%%%%%%%%%%
\newpage

\section{Tables}

\begin{table*}[!!h]
\centering
\caption{Notation used in the model}
\begin{tabular} {l l l}
\hline\hline
Notation &  Description & Units\\ [0.5ex]
\hline
$W$ & Local energy per unit area &  pN/nm\\
$H$ & Mean curvature of the membrane & nm$^{-1}$\\
$K$ & Gaussian curvature of the membrane & nm$^{-2}$\\
$C$ & Prescribed spontaneous curvature & nm$^{-1}$ \\
$\theta^{\alpha}$ & Parameters describing the surface, $\alpha \in \{1,2\}$\\
$A_{\mathrm{coat}}$ & Area covered by the coat & nm$^2$ \\
${\bf r}$ & Position vector & \\
${\bf n}$ & Normal to the membrane surface &  unit vector\\
$\textbf{a}_\alpha$ & Basis vectors describing the tangent plane,  $\alpha \in \{1,2\}$\\
$\gamma$ & Lagrange multiplier for the incompressibility constraint & pN/nm\\
$\lambda$ & Membrane tension, $-(W+\gamma)$  & pN/nm\\
$p$ & Pressure difference across the membrane & pN/nm$^2$\\
$\textbf{f}$ & Applied force per unit area & pN/nm$^2$ \\
$k$ & Bending modulus  & pN$\cdot$nm\\
$\bar{k}$ & Gaussian modulus & pN$\cdot$nm\\
$s$ & Arc-length & nm\\
$S$ & Total arc-length of the membrane patch & nm\\
$\theta$ & Azimuthal angle & \\
$r$ & Radial distance & nm\\
$z$ & Elevation from base plane & nm\\
$\mathbf{e}_{r}$ & Radial basis vector & unit vector \\
$\mathbf{e}_{\theta}$ & Azimuthal basis vector & unit vector \\
$\mathbf{k}$ & Altitudinal basis vector & unit vector \\
$\mathbf{a}_s$ & Tangent to the membrane surface in the radial direction & unit vector\\
$\psi$ & Angle between $\mathbf{e}_{r}$ and $\mathbf{a}_s$ &\\
\hline
\end{tabular}
\label{table:notation1}
\end{table*}

\begin{table*}[!!h]
\centering
\caption{Notation used in the model (continued)}
\begin{tabular} {l l l}
\hline\hline
Notation &  Description & Units\\ [0.5ex]
\hline
$\kappa_{\nu}$ & Tangential curvature & nm$^{-1}$\\
$\kappa_{t}$ & Transverse curvature & nm$^{-1}$\\
$L$ & Shape equation variable & nm$^{-1}$\\
$Z_p$ & Prescribed displacement at the pole & nm\\
$H_p$ & Prescribed mean curvature at the pole & nm$^{-1}$\\
$t$ & Dimensionless arc-length &\\
$x$ & Dimensionless radial distance &\\
$y$ & Dimensionless height &\\
$h$ & Dimensionless mean curvature &\\
$c$ & Dimensionless spontaneous curvature &\\
$l$ & Dimensionless $L$ &\\
$\widetilde{\lambda}$ & Dimensionless membrane tension &\\
$\widetilde{p}$ & Dimensionless transmembrane pressure &\\
$\widetilde{\mathbf{f}}$ & Dimensionless force per unit area &\\
$\widetilde{k}$ & Dimensionless bending rigidity &\\
$T$ & Total dimensionless arc-length &\\
$y_p$ & Dimensionless prescribed pole displacement &\\
$H_p$ & Dimensionless prescribed pole mean curvature &\\
$a$ & Membrane area & nm$^{2}$\\
$A$ & Total area of membrane patch & nm$^{2}$\\
$\alpha$ & Dimensionless membrane area & \\
$\alpha_{max}$ & Total dimensionless area & \\
\hline
\end{tabular}
\label{table:notation2}
\end{table*}

\begin{table*}[!!h]
\centering
\caption{Parameters used in the model}
\begin{tabular} {l l l l}
\hline\hline
Parameter  & Significance & Value & Reference  \\ [0.5ex]
\hline
$\lambda_0$ & Membrane tension range & $10^{-4} - 1$  pN/nm & \cite{Dai1998,Stachowiak2013,Sens2015}\\%,Lipowsky2012,Pecreaux2004,Faris2009} \\
$k$ & Bending rigidity of bare membrane & $320$ pN $\cdot$ nm & \cite{Stachowiak2013} \\ %\cite{Lipowsky1995}\\
$C_0$ & Preferred curvature of coat & $1/50$ nm$^{-1}$ & \cite{Stachowiak2013,Miller2015}\\
$R_0$ & Non-dimensionalization length & $20$ nm \\
\hline
\end{tabular}
\label{table:parameters}
\end{table*}

\clearpage
%\printbibliography

%%%%%%%%%%%%%%%%%%%%%%%%%%%%%%%%%%%%%%%%%%%

\newpage

\section{Supplementary Figures}

%%%%%%%%%%%%%%%%%%%%%%%%%%%%%%%%%%%%%%%%%%

%%%%%%%%%%%%%%%%%%%%%%%%%%%%%%%%%%%%%%%%%%

%%%%%%%%%%%%%%%%%%%%%%%%%%%%%%%%%%%%%%%%%%

%%%%%%%%%%%%%%%%%%%%%%%%%%%%%%%%%%%%%%%%%%%

%%%%%%%%%%%%%%%%%%%%%%%%%%%%%%%%%%%%%%%%%%

\begin{figure}
\centering

\includegraphics[width=0.95\linewidth, height = 0.7\textheight, keepaspectratio]{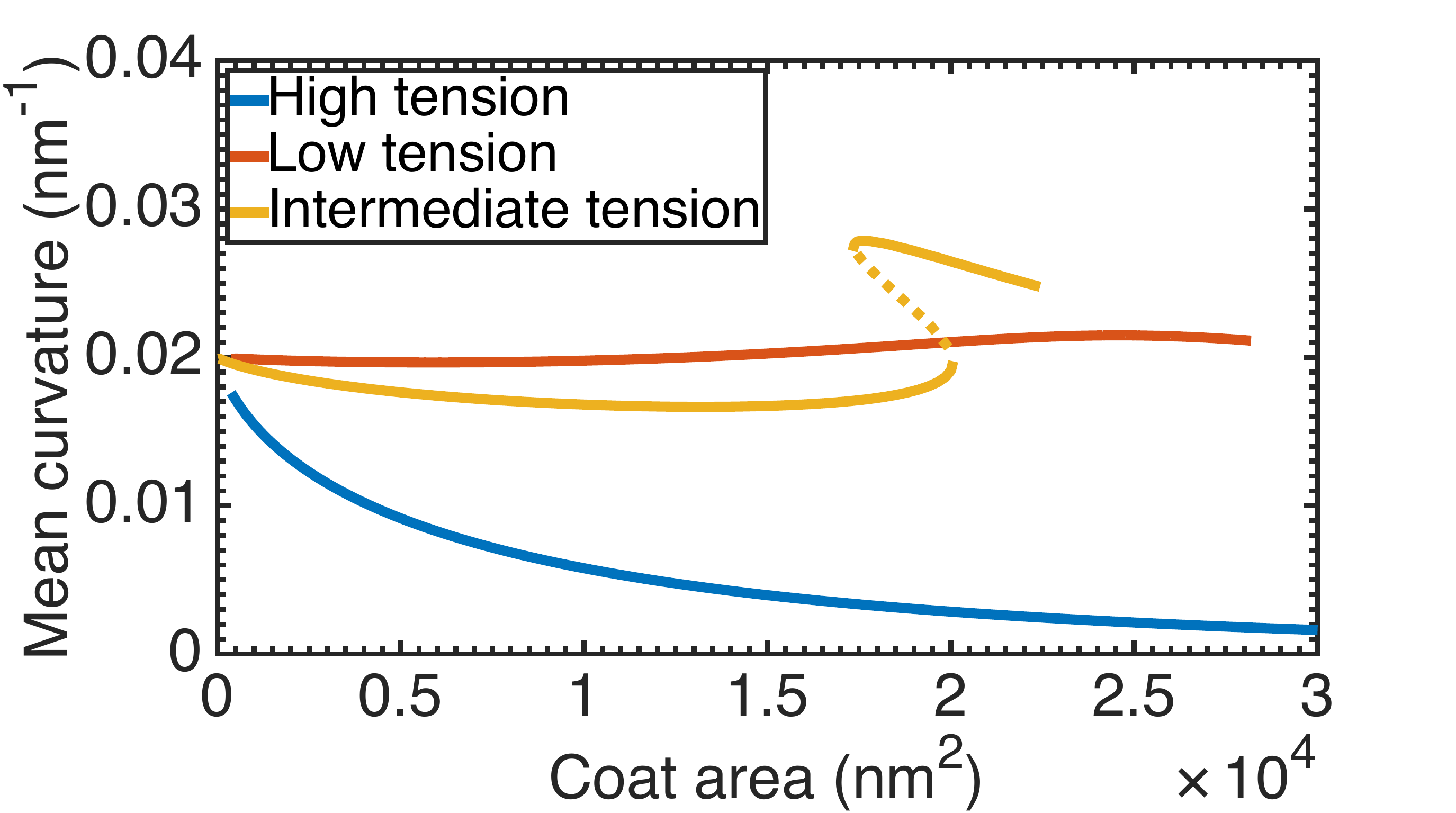}

\caption{Mean curvature at the bud tip as a function of the area of the coat for the three different membrane tension cases. \textcolor{myBlue}{High membrane tension}, $\lambda_0 = 0.2 \,\mathrm{pN/nm}$: The mean curvature at the bud tip drops to nearly $0 \,\mathrm{nm^{-1}}$ as the size of the coat increases and the membrane stays essentially flat at the center of the patch (Figure \ref{fig:tensionComp}A-C). \textcolor{myRed}{Low membrane tension}, $\lambda_0 = 0.002 \,\mathrm{pN/nm}$: The mean curvature at the bud tip remains at approximately $0.02 \,\mathrm{nm^{-1}}$ as the size of the coat increases and the membrane adopts the spontaneous curvature of the coat (Figure \ref{fig:tensionComp}D-F). \textcolor{myYellow}{Intermediate membrane tension}, $\lambda_0 = 0.002 \,\mathrm{pN/nm}$: Reproduced from Figure \ref{fig:snapthrough}B. The mean curvature at the bud tip is lower for open buds (lower solution branch) relative to the low tension case, indicating that tension is inhibiting curvature generation by the coat. In contrast, the curvature is higher in the closed buds (upper solution branch) relative to closed buds in the low tension case, showing that membrane tension serves to shrink the size (and hence increase the curvature) of closed buds.}

\label{fig:HvsA0}

\end{figure}

%%%%%%%%%%%%%%%%%%%%%%%%%%%%%%%%%%%%%%%%%%

\begin{figure}[h]
\centering
\includegraphics[width=\linewidth, height = 0.7\textheight, keepaspectratio]{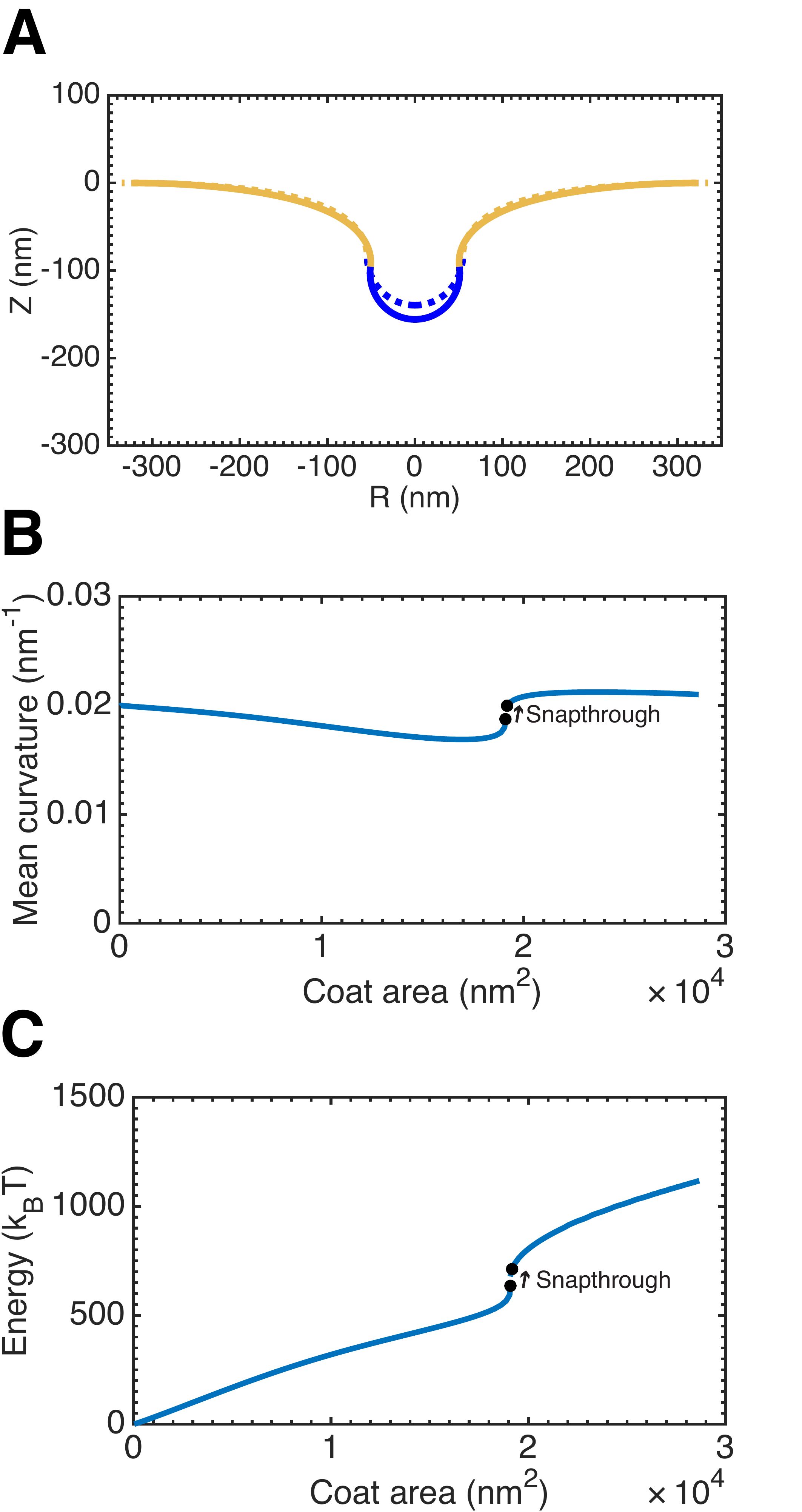}

\caption{
A snapthrough instability is present with increased coat stiffness when progressively increasing the coat area at intermediate membrane tension, $\lambda_0 = 0.02 \,\textrm{pN}/\textrm{nm}$. The coat stiffness was increased to $k_{\mathrm{coat}} = 2400 \,\textrm{pN}\cdot\textrm{nm} \approx 300 \,k_{\mathrm{B}} T$ \cite{Jin2006}, while maintaining uncoated membrane stiffness at $k_{\mathrm{coat}} = 320 \,\textrm{pN}\cdot\textrm{nm} \approx 80 \,k_{\mathrm{B}} T$ . \textbf{(A)} Membrane profile showing the morphology before (dashed line, $A_{\mathrm{coat}}= 19,083 \,\mathrm{nm}^2$) and after (solid line, $A_{\mathrm{coat}}= 19,171 \,\mathrm{nm}^2$) a small addition of area to the coat. \textbf{(B-C)} Mean curvature at the tip of the bud and energy necessary to deform the membrane as a function of the area of the coat. A snapthrough instability is present with a stiffer coat, though the energy difference between the U- and $\Omega$-shaped pits is only $\approx 75 \,k_{\mathrm{B}} T$ and the solution branches overlap very slightly.
}

\label{fig:stiffCoat}

\end{figure}

%%%%%%%%%%%%%%%%%%%%%%%%%%%%%%%%%%%%%%

\begin{figure}

\includegraphics[width=0.95\linewidth, height = 0.7\textheight, keepaspectratio]{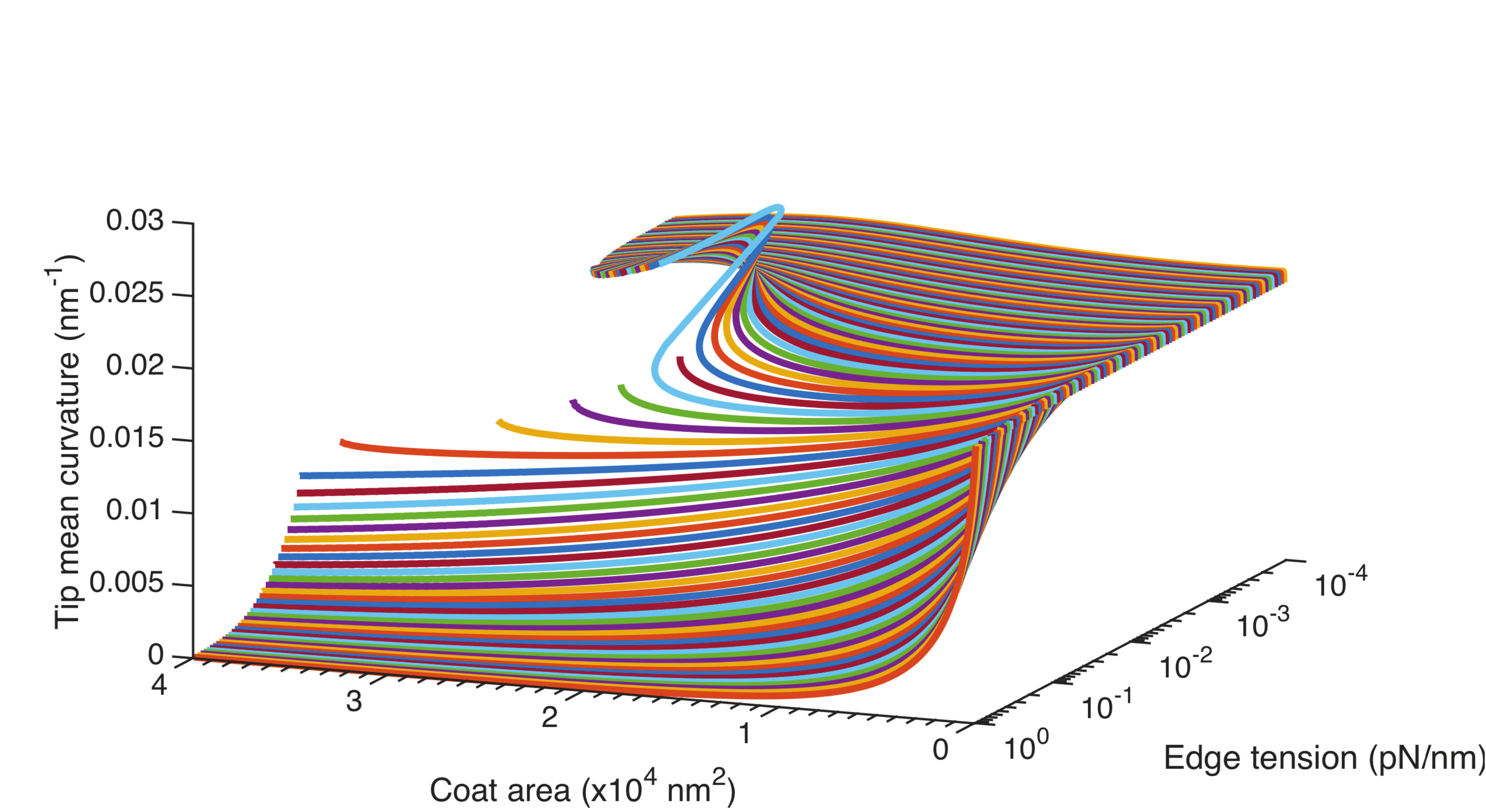}

\caption{Cusp catastrope surface. The mean curvature at the tip of the bud is plotted as a function of the membrane tension and coat area. Membrane tension is on a log scale. Three regimes exist: 1) Low tension: The membrane smoothly evolves from flat to a closed bud. The tip mean curvature remains nearly constant at the preferred curvature of the coat. 2) High tension: The membrane remains nearly flat as the coat area increases. The tip mean curvature goes to zero as the size of the coat increases. 3) Intermediate tension: A snapthrough instability in the tip mean curvature exists after a bifurcation point is reached.}

\label{fig:cusp}

\end{figure}

%%%%%%%%%%%%%%%%%%%%%%%%%%%%%%%%%%%

\begin{figure}

\centering
\includegraphics[width=0.95\linewidth, height = 0.7\textheight, keepaspectratio]{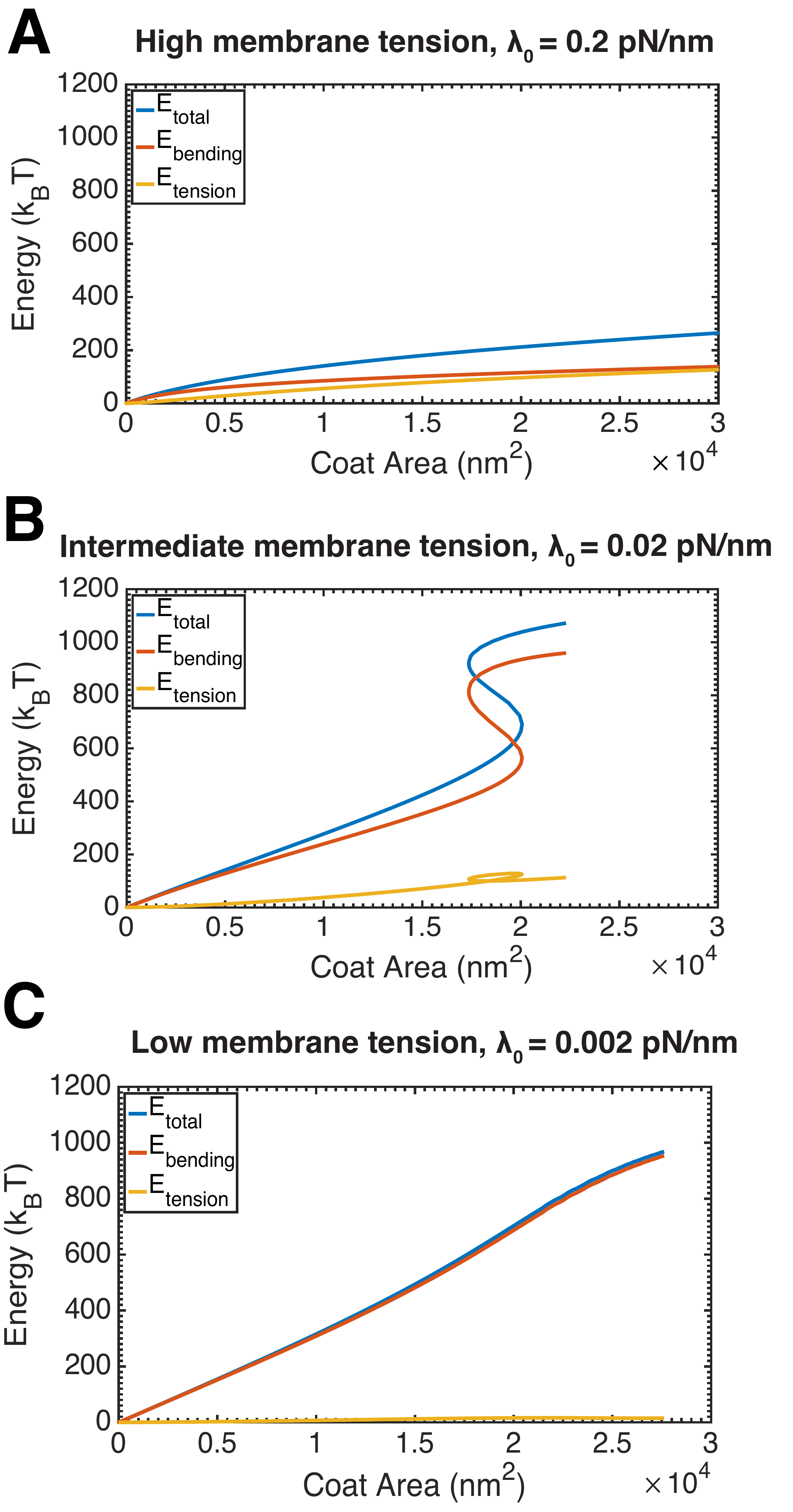}

\caption{Energy to deform the membrane as a function of the coat area. \textbf{(A)} High membrane tension, $\lambda_0 = 0.2 \, \textrm{pN}/\textrm{nm}$: There is no substantial deformation of the membrane at high tensions by the coat, and so the total work done to deform the membrane is relatively low. The work done against bending rigidity and against tension are of the same order of magnitude. \textbf{(B)} Intermediate membrane tension, $\lambda_0 = 0.2 \, \textrm{pN}/\textrm{nm}$: The main contribution to the deformation energy is work done against bending rigidity. Importantly, the energy barrier between open and closed bud morphologies is still present in the work done against bending rigidity and is not just a consequence of work done against the membrane tension. \textbf{(C)} Low membrane tension, $\lambda_0 = 0.2 \, \textrm{pN}/\textrm{nm}$: Nearly all of the work done is against bending rigidity.}

\label{fig:eVsCarea}

\end{figure}

%%%%%%%%%%%%%%%%%%%%%%%%%%%%%%%%%%%%%%%%%%

%%%%%%%%%%%%%%%%%%%%%%%%%%%%%%%%%%%%

\end{document}